\documentclass[extra,onecolumn]{gji}
\usepackage{amsmath,amsfonts}
\usepackage{graphicx}
\begin{document}
\title[Radiative transfer of body and surface waves]{A scalar radiative transfer model including the coupling between surface and body waves}
\author[L. Margerin et al.]{Ludovic Margerin$^1$, Andres Bajaras$^2$, Michel Campillo$^2$ \\ $^1$ Institut de Recherche en Astrophysique et Plan\'etologie,   Observatoire Midi-Pyr\'en\'ees,  \\ Universit\'e Paul Sabatier, C.N.R.S., C.N.E.S.,  14 Avenue Edouard Belin, 31400 Toulouse,  France.
 \\$^2$  Institut des Sciences de la Terre, Observatoire des Sciences de l'Univers de Grenoble, \\ Université Grenoble Alpes, C.N.R.S., I.R.D., CS 40700,
38058 GRENOBLE Cedex 9, France }

\date{\today}
\maketitle
\begin{abstract}
To describe the energy transport in the seismic coda, we introduce a system of radiative transfer equations for coupled surface and body waves in a scalar approximation. Our model is based on the Helmholtz equation in a half-space geometry with mixed boundary conditions. In this model, Green's function can be represented as a sum of body waves and surface waves, which mimics the situation on Earth. In a first step, we  study the single-scattering problem for point-like objects in the Born approximation. Using the assumption that the phase of body waves is randomized by surface reflection or by interaction with the scatterers, we show that it becomes possible to define, in the usual manner, the cross-sections for surface-to-body and body-to-surface scattering. Adopting the independent scattering approximation, we then define the scattering mean free paths of body and surface waves including the coupling between the two types of waves. Using a phenomenological approach, we then derive a set of coupled transport equations satisfied by the specific energy density of surface and body waves in a medium containing a homogeneous distribution of point scatterers. In our model, the scattering mean free path of body waves is depth dependent as a consequence of the body-to-surface coupling. We demonstrate that an equipartition between surface and body waves is established  at long lapse-time, with a ratio which is predicted by usual mode counting arguments. We derive a diffusion approximation from the set of transport equations and show that the diffusivity is both anisotropic and depth dependent. The physical origin of the two properties is discussed. Finally, we present Monte-Carlo solutions of the transport equations which illustrate the convergence towards equipartition at long lapse-time as well as the importance of the coupling between surface and body waves in the generation of coda waves.
\end{abstract}
\section{Introduction}
In seismology, Radiative Transfer (RT) has been used for more than three decades to characterize the scattering and absorption properties of Earth's crust \citep[see for instance][]{fehler1992,hoshiba1993,carcole2010,eulenfeld2017}.  Since its introduction by \citet{wu1985} for scalar waves in the stationary regime, RT has been considerably improved to bring it in closer agreement with real-world applications. In particular, the coupling between shear and compressional waves was developed in a series of papers by  \citet{weaver1990,turner1994,zeng1993,sato1994,ryzhik1996}. The model of \citet{sato1994} was applied to data from an active experiment by \citet{yamamoto2010} and showed impressive agreement between observations and elastic RT theory. For comprehensive introductions to RT, the reader is referred to the review chapter by \citet{margerin2005} or the monograph of \citet{sato2012}.

Parallel to the physical and mathematical developments of the theory, more and more realistic Monte-Carlo simulations of the transport process were developed over the years. This includes, for example, the treatment of non-isotropic scattering \citep{abubakirov1990,hoshiba1995,gusev1996,jing2014,sato2018}, velocity and heterogeneity stratification \citep{hoshiba1997,margerin1998,yoshimoto2000}, coupling between shear and compressional waves \citep{margerin2000,przybilla2009},  laterally varying velocity and scattering structures \citep{sanborn2017}.
With the growth of computational power, Monte-Carlo simulations opened up new venues for the application of RT in seismology:  imaging of deep Earth heterogeneity \citep{margerin2003b,shearer2004,mancinelli2013,mancinelli2016frequency}, mapping of the depth-dependent scattering and absorption structure of the lithosphere \citep{mancinelli2016constraints,takeuchi2017}, modeling of propagation anomalies in the crust \citep{sens2009,sanborn2018}, P-to-S conversions in the teleseismic coda \citep{gaebler2015}, to cite a few examples only.   
 
 Recently, RT has also been applied to the computation of sensitivity kernels for time-lapse imaging methods such as coda wave interferometry \citep[see e.g.][]{poupinet1984,snieder2006,poupinet2008}. Coda Wave Interferometry (CWI) exploits tiny changes of waveforms in the coda to map the temporal variations of seismic properties in 3-D. The mapping relies on the key concept of sensitivity kernels, which, in the framework of CWI, were introduced by \citet{pacheco2005} in the diffusion regime and \citet{pacheco2006} in the single-scattering regime. These kernels take the form of spatio-temporal convolutions of the mean intensity in the coda. It was later pointed out by \citet{margerin2016} that an accurate computation of traveltime sensitivity kernels, valid for an arbitrary scattering order and an arbitrary spatial position, requires the knowledge of the angular distribution of energy fluxes in the coda. These fluxes, or specific intensities, are directly predicted by the radiative transfer model, which makes it attractive for imaging applications.  
 
 In noise-based monitoring \citep{wegler2007} -also known as Passive Image Interferometry (PII)- the  virtual sources and receivers are located at the surface of the medium so that the early coda is expected to contain a significant proportion of Rayleigh waves. At longer lapse-time, the surface waves couple with body waves and the coda eventually reaches an equipartition regime when all the propagative surface and body wave modes are excited to equal energy \citep{weaver1982,hennino2001}. Because the volumes explored by surface and body waves are  significantly different, the knowledge of the composition of the coda wavefield at a given lapse-time in the coda is key to locate accurately the changes at depth in the crust. 
 
 \citet{obermann2013,obermann2016} proposed to express the sensitivity of coda waves as a linear combination of the sensitivity of surface and body waves, whose kernels are computed from scalar RT theory in 2-D and 3-D, respectively. The relative contribution of the 2-D and 3-D sensitivity kernels at a given lapse-time in the coda is determined by fitting the traveltime shift predicted by the theory against full wavefield numerical simulations in scattering media, where the background seismic velocity is perturbed in a fine layer at a given depth. This method has been validated in the case of 1-D perturbations through numerical tests and has the advantage of modeling exactly the complex coupling between surface and body waves in heterogeneous media. Furthermore, it can easily incorporate realistic topographies, which is important for the monitoring of volcanoes.  The main drawbacks of the approach of \citet{obermann2016} are the numerical cost and the fact that it requires a good knowledge of the scattering properties of the medium, which have to be determined by other methods such as MLTWA \citep{fehler1992,hoshiba1993}.
 
 This brief overview illustrates that CWI and PII  would benefit from a formulation of RT theory which incorporates the coupling between surface and body waves in a self-consistent way. In the case of a slab bounded by two free surfaces, \cite{tregoures2002} derived from first principles a quasi 2-D RT equation where the wavefield is expanded onto a basis of Rayleigh, Lamb and Love eigenmodes. Thanks to the normal mode decomposition, this model incorporates the boundary conditions at the level of the wave equation. The energy exchange between surface and body waves is treated by normal mode coupling in the Born approximation. A notable advantage of this formulation is the capacity to predict directly the energy decay in the coda and its parttioning onto different components. The two main limitations for seismological applications are the slab geometry, which may not always be realistic and the fact that the disorder should be weak, i.e., the mean free time should be large compared to the  vertical transit time of the waves through the slab.
 
  \citet{zeng2006} proposed a system of coupled integral equations to describe the exchange of energy between surface waves and body waves in the seismic coda. The formalism used by the author is interesting and bears some similarities with the one developed in this work, although we formulate the theory in integro-differential form. A blind spot in the work of \citet{zeng2006}  is the coupling between surface and body waves, which  is introduced in an entirely phenomenological way and differs significantly from our findings.  A very promising investigation of the energy exchange between surface and body waves on the basis of the elastodynamic equations  in a half-space geometry was performed by \citet{maeda2008}. Using the Born approximation, these authors calculated the scattering coefficients between all possible modes of propagation in a medium containing random inhomogeneities. The main limitation of their theory comes from the fact that the conversion from body to surface waves is quantified by a non-dimensional coefficient, which makes it difficult to extend their results beyond the single-scattering regime. The authors argue that the absence of a characteristic scale-length for body-to-surface attenuation is a consequence of the fact that all conversions occur in approximately one Rayleigh wavelength in the vicinity of the surface.
 
 In this work, we revisit the problem of coupling surface and body waves in a RT framework using an approach similar to the one of \citet{maeda2008}. For simplicity, we limit our investigations to a scalar model based on the Helmholtz equation with an impedance (or mixed) boundary condition in a half-space geometry. To make the presentation self-contained, we review the most important features of this particular wave equation. Specifically, we recall that the modes of propagation are composed of body waves, and surface waves whose penetration depth depends on the impedance condition only. Hence, our model mimics the situation on Earth while minimizing the mathematical complexity. We then introduce a simple point-scattering model and study its properties in the Born approximation. Using the additional assumption that the surface reflexion randomizes the phase of the reflected wave, we are able to derive simple expressions for the scattering mean free path of both body and surface waves including the coupling between the two. We elaborate on this result to establish a set of two coupled RT equations satisfied by the specific energy density of surface and body waves using a phenomenological approach. Some consequences of our simple theory are explored, in particular the establishment of a diffusion and equipartition regime. Monte-Carlo simulations show the potential of the approach to model the transport of energy in the seismic coda from single-scattering to diffusion.
\section{Scalar wave equation model with surface and body waves}
In this section, we present the basic ingredients of our scalar model based on the Helmholtz equation. We describe how an appropriate modification of boundary conditions at the surface of a half-space gives rise to the presence of a surface wave. We subsequently present an expression of the Green's function and its asymptotic approximation. The concept of density of states, important for later developments, is recalled. For a thorough treatment of the mathematical foundations of our model, the interested reader is refered to the monograph of \citet{hein2010}.
\subsection{Equation of motion}
We consider a 3-D version of the membrane vibration equation in a half-space geometry:
\begin{equation}
  \left( \rho \partial_{tt}   -T \Delta \right) u(t,\mathbf{R}) =0   \label{membrane}
\end{equation}
where $t$ is time and $\mathbf{R}$ is the position vector. It may be further decomposed as  $\mathbf{R} =\mathbf{r} + z\mathbf{\hat{z}}$ ($z\geq0$)  with $\mathbf{r} =x\mathbf{\hat{x}} +y \mathbf{\hat{y}} $ and $(\mathbf{\hat{x}}, \mathbf{\hat{y}}, \mathbf{\hat{z}})$ denotes a Cartesian system. In Eq. (\ref{membrane}) $u$, $\rho$ and $T$  may be thought of as the displacement,  the density and the elastic constant of the medium, respectively. The wave Eq. (\ref{membrane})  is supplemented with the boundary conditions:
\begin{equation}
\begin{split}
   \left. (\partial_z  +\alpha) u(t,\mathbf{R})\right|_{z=0}  =0  \label{BC}   \\
                 + \text{radiation condition at } \infty   
  \end{split}
\end{equation}
The case of interest to us corresponds to $\alpha >0$, i.e. when, as recalled below, the boundary can support a surface wave.
Equation (\ref{membrane}) can be derived by applying Hamilton's principle to the following Lagrangian density:
\begin{equation}
L =  \dfrac{1}{2} \left[ \rho (\partial_t u(t,\mathbf{R}))^2 - T ( \nabla u(t,\mathbf{R}))^2  + \alpha T u(t,\mathbf{R})^2 \delta(z) \right]  \label{lagrangian}
\end{equation}
Thanks to the last term of the Lagrangian (\ref{lagrangian}), which corresponds to a negative elastic potential energy stored at the surface $z=0$, the first B.C. in Eq. (\ref{BC}) becomes natural in the sense of variational principles.  To make the presentation self-contained, we explain in Appendix \ref{variational} the origin of the delta function  in Eq. (\ref{lagrangian}) in the simple case of a finite string with mixed boundary conditions at one end and free boundary conditions at the other end.

In the case of  a harmonic time dependence  $u \propto e^{-i\omega t}$ ($\omega >0$), the vibrations are governed by Helmholtz Eq.:
\begin{equation}
 \Delta u(\mathbf{R}) + \dfrac{\omega^2}{c^2} u(\mathbf{R})= 0  \label{helmholtz}
\end{equation}
with $c=\sqrt{T/\rho}$ the speed of propagation of the waves in the bulk of the medium. Eq. (\ref{helmholtz}) is complemented with the mixed boundary condition $\partial_z u + \alpha u=0$ at $z=0$ and an outgoing wave condition at infinity.
From (\ref{lagrangian}), we can deduce the energy flux density vector $\mathbf{J}$ and the energy density $w$ using the concept of stress-energy tensor \citep{morse1986}.
For harmonic motions, their average value over a period can be expressed as:
\begin{align}
\mathbf{J} = & -\dfrac{T}{2} \text{Re} \left\{ i \omega u^* \nabla u   \right\}  \label{J} \\
 w   =   &  \dfrac{1}{4} \left\{ \rho \omega^2 |u|^2  +  T | \nabla u|^2 - \alpha T |u|^2 \delta(z) \right\}  \label{w}
\end{align}
In the following section, we recall the consequences of mixed boundary conditions on the Helmoltz Eq., in particular the fact that it gives rise to a surface wave mode.
\subsection{Eigenfunctions and Green's function}
Due to the translational invariance of the medium, we look for eigen-solutions of Eq. (\ref{helmholtz}) in the form $u=\psi(z) e^{i  \mathbf{k}_{\parallel} \cdot \mathbf{r} }$
with  $ \mathbf{k}_{\parallel} = (k_x,k_y,0)$.  This leads to a self-adjoint eigenvalue problem in the $z$ variable only.
For $\alpha>0$, part of the spectrum is discrete with eigenfunction :
\begin{equation}
u_s(\mathbf{r},z) = \sqrt{2\alpha}e^{-\alpha z} \dfrac{e^{i \mathbf{k}_{\parallel} \cdot \mathbf{r} }}{2\pi} \label{swm}
\end{equation}
with $\mathbf{k}_{\parallel} \cdot  \mathbf{k}_{\parallel} - \alpha^2 = \dfrac{\omega^2}{c^2}$. The rest of the spectrum forms a continuum of body waves with normalized eigenfunctions:
\begin{equation}
 u_{b}(\mathbf{r},z) =  \dfrac{1}{(2\pi)^{3/2}} (e^{-iq z} + r(q) e^{iqz}) e^{i \mathbf{k}_{\parallel} \cdot \mathbf{r} }  , q\geq 0  \label{bwm}
\end{equation}
with $q^2 + \mathbf{k}_{\parallel} \cdot  \mathbf{k}_{\parallel} =  \dfrac{\omega^2}{c^2}$  and:
\begin{equation}
 r(q) = \dfrac{q + i\alpha}{q - i\alpha}
\end{equation}
Note the relations:  (1) $q = (\omega \cos j)/c = k \cos j$   with $j$ the incidence angle of the body wave and (2) $|r(q)|^2 =1$, i.e., there is total reflection at the surface.  For later reference, we introduce a specific notation for the vertical eigenfunction of body waves:
\begin{equation}
\psi_b(\mathbf{\hat{n}},z) = (e^{-i k z\mathbf{\hat{n}} \cdot \mathbf{\hat{z}}} + r(k \mathbf{\hat{n}} \cdot \mathbf{\hat{z}}) e^{i k z \mathbf{\hat{n}} \cdot \mathbf{\hat{z}}} ) \label{psib}
\end{equation}
with $\mathbf{\hat{n}} \cdot \mathbf{\hat{z}} =\cos j$. Note that throughout the paper, we use a hat to denote a unit vector.
The  surface waves (\ref{swm}) and body waves (\ref{bwm}) are normalized and orthogonal in the sense of the scalar product $\langle u | v \rangle = \int_{\mathbb{R}^3_{+}} u(\mathbf{R})^* v(\mathbf{R})  d^3R$,  where $u$ and $v$ are arbitrary square integrable functions. 
The surface wave phase velocity $c_{\phi}$ is given by: 
\begin{equation}
 c_{\phi} = \dfrac{c}{\sqrt{1 + \dfrac{c^2 \alpha^2}{\omega^2}  }} \label{cphi}
\end{equation}
and is always smaller than the bulk wave velocity $c$.
The group velocity may be obtained in two different manners, namely,
(1) using the  classical formula based on the interference of a wave packet:
 \begin{equation}
      v_g = \dfrac{d\omega}{dk} = \dfrac{c^2}{c_{\phi}} = c \sqrt{1 + \dfrac{c^2 \alpha^2}{\omega^2}  }  \label{domdk}
 \end{equation}
 and (2) using the principle of energy conservation:   
 \begin{equation}
   \mathbf{v}^E_s = \dfrac{\langle \mathbf{J} \rangle}{\langle w \rangle} =  v_g \mathbf{\hat{k}}_{\parallel}  \label{evel}
 \end{equation}
 where the brackets indicate  an integration over the whole depth range.
 
 \begin{figure}
 \centerline{\includegraphics[width=0.6\linewidth]{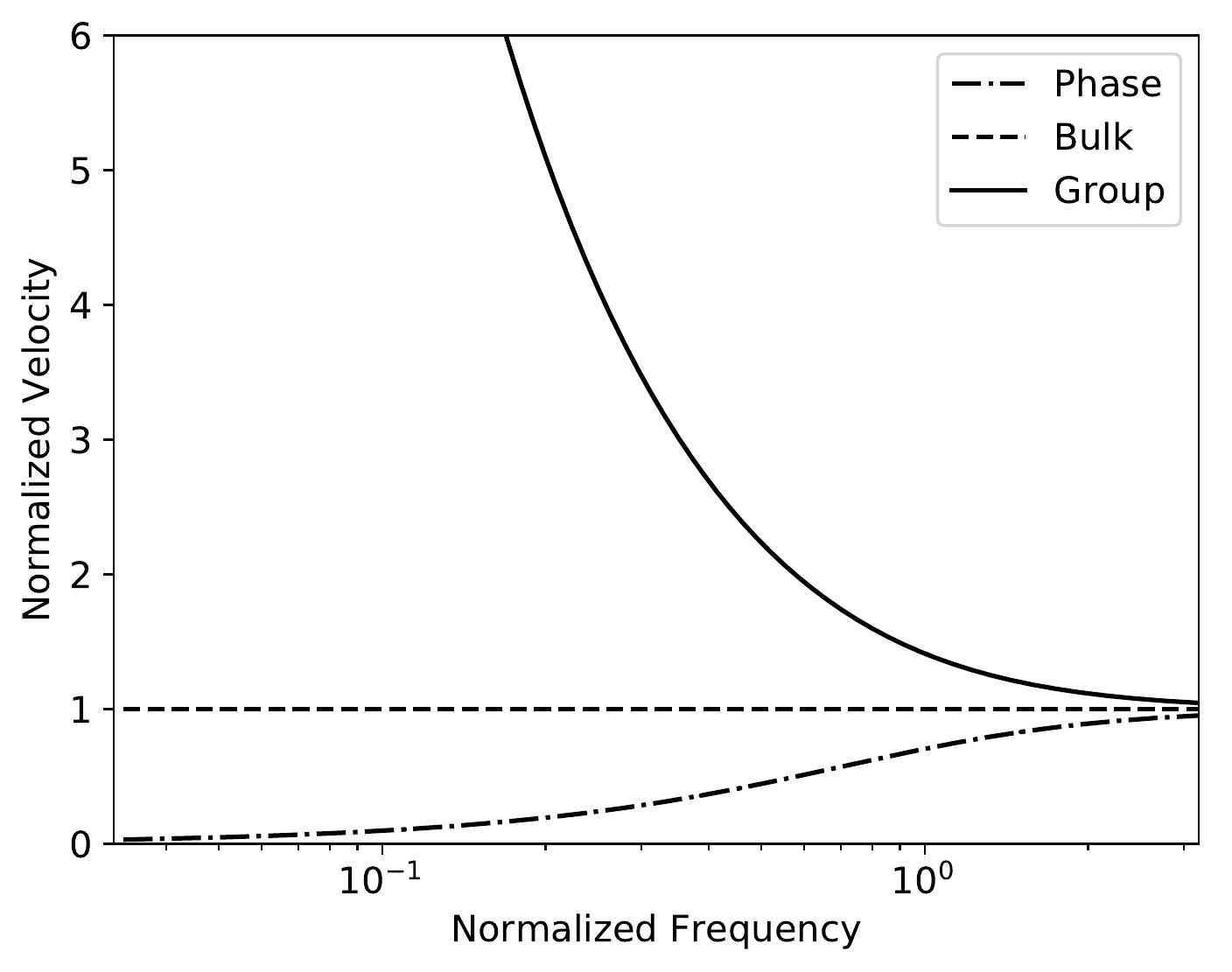}}
 \caption{Dispersion of surface waves in a half-space with mixed boundary conditions. On the horizontal axis, the normalized frequency is defined as $\omega \alpha/c$. On the vertical axis, the velocity is normalized by the speed of  body waves $c$.} \label{dispersion}
 \end{figure}

 The dispersion properties of the surface wave in our scalar model are illustrated in Figure \ref{dispersion}. As is evident from  Eqs (\ref{cphi})-(\ref{domdk}), the group velocity is always faster than both the phase and body wave velocity. In the high-frequency limit, the phase and group velocity tend to the common value $c$.
Using definition (\ref{evel}) it is possible to define the energy velocity of a body wave eigenmode (see Eq. \ref{bwm}):
 \begin{equation}
   \mathbf{v}^E_b =  \lim_{h \to \infty} \dfrac{\langle \mathbf{J} \rangle_h}{\langle w \rangle_h} =  c \sin j \mathbf{\hat{k}}_{\parallel} , \label{ve}
 \end{equation}
 where  $\langle \rangle_h$   denotes an integration from the surface to depth $h$.  The depth averaging smoothes out the oscillations of $\mathbf{J}$ and $w$ caused by the interference between the incident and reflected amplitudes.  The passage to the limit is necessary because the integrals over depth diverge.   Eq. (\ref{ve}) can be interpreted as follows.  In the full-space case, the current vector of a single unit-amplitude plane wave with wavevector $\mathbf{k} = k (\cos j \mathbf{\hat{z}}  + \sin j \mathbf{\hat{k}}_{\parallel}) $ is given by $\mathbf{J} = \rho \omega c^2 \mathbf{k}/2$ and  carries an energy density $w = \rho \omega^2/2$. If we define $\mathbf{k}_r$ as the mirror image of $\mathbf{k}$ across the plane $z=0$ and consider the sum of  the current vector of two plane waves with wavevectors $\mathbf{k}$ and  $\mathbf{k}_r$ we obtain $\mathbf{J} + \mathbf{J}_r = \rho \omega^2 c \sin j \mathbf{\hat{k}}_{\parallel}$. After normalization by the sum of energy densities, the result (\ref{ve}) is recovered. 
In other words, on average, the energy transported by a body wave mode is simply the sum of the energies transported by the incident and reflected waves, as if the two  were independent.  

Using the eigenmodes (\ref{swm}) and (\ref{bwm}), one may obtain an exact representation of the Green's function of Helmholtz Eq. with mixed BC in the form:
 \begin{equation}
 \begin{split}
 G(\mathbf{r},z,z_0) = & \dfrac{1}{(2\pi)^3} \int^{+\infty}_0 dq \int_{\mathbb{R}^2} \dfrac{e^{i \mathbf{k}_{\parallel} \cdot \mathbf{r} } (e^{-i q z}  +r(q) e^{iq z}) (e^{-i q z_0}  +r(q) e^{iq z_0})^* }{k^2 - k_{\parallel}^2 -q^2 +i \epsilon } d^2k_{\parallel} \\
           &+\dfrac{2 \alpha e^{-\alpha(z+z_0)}}{(2 \pi)^2}    \int_{\mathbb{R}^2} \dfrac{e^{i \mathbf{k}_{\parallel} \cdot \mathbf{r} } }{k^2 +\alpha^2 -k_{\parallel}^2 +i\epsilon}    d^2k_{\parallel} 
        \end{split} , \label{gex}
 \end{equation}
 where $z_0$ denotes the source depth, $\epsilon$ is a small positive number which guarantees the convergence of the integrals  and the star $^*$ denotes complex conjugation. In Eq. (\ref{gex}) the first (resp. second) line represents the body wave (resp. surface wave) contribution.  As shown in Appendix \ref{gfderiv}, the surface wave term can be computed analytically in terms of Hankel functions.
  The following far-field approximation of the Green's function of the Helmholtz Eq. (\ref{helmholtz}) can be obtained using the stationary phase approximation for the body wave term:
 \begin{equation}
    G(\mathbf{r},z,z_0)  =   -\dfrac{e^{i k R}}{4 \pi R} \psi_b(\mathbf{\hat{R}},z_0) - \dfrac{\alpha e^{-\alpha(z + z_0) + i k_s r +i \pi/4 } }{\sqrt{2 \pi k_s r  }}  \label{gff}  
  \end{equation}
  with $k_s = \omega / c_{\phi}$,  $ \mathbf {\hat{R}}= \mathbf{R}/R$ and $\mathbf{R} = \mathbf{r} + z \mathbf{\hat{z}}$. The expansion (\ref{gff}) is performed with respect to the midpoint of the source point and its mirror image by the surface $z=0$. 
 The $z$ dependence of the first term is simply given by the body wave eigenfunction (\ref{bwm}).  For further computational details, the reader may consult Appendix \ref{gfderiv}.
 
 \subsection{Source radiation and density of states}
 \begin{figure}
 \centerline{\includegraphics[width=0.7\linewidth]{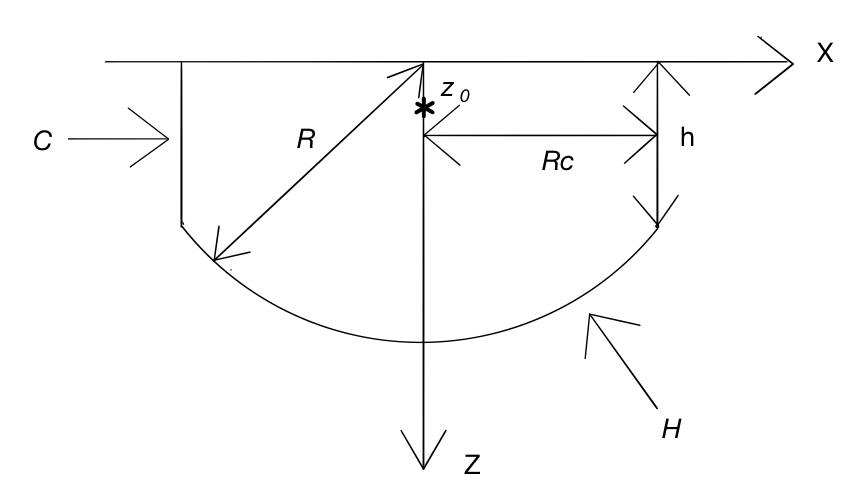}}
 \caption{Geometry of the surface employed to compute the energy radiation of a point source located at $(0,0,z_0)$. $C$ is a cylindrical surface of radius $R_c$ and height $h$. $H$ is a hemispherical surface of radius $R$. } \label{figch}
 \end{figure}
   We now compute the energy radiated by a point source located at $(0,0,z_0)$. To do so, we introduce a cylindrical surface $C$ of radius $R_c$ extending from the free surface to a depth $h$ greater than $z_0$ and large compared to $1/\alpha$.
   We close this surface with a hemispherical cap $H$ of radius $R$ centered at the surface point $(0,0,0)$. The geometry is schematically depicted in Figure \ref{figch}. The energy flux vector (\ref{J}) of the radiated field  contains terms that are purely surface, purely bulk and cross-terms. The contribution of surface waves to the flux across the  hemispherical surface is negligible (the error made is exponentially small). The contribution of body waves to the flux across the lateral cylindrical surface is also negligible because this surface subtends a solid angle which goes to $0$ as $R_c$ increases. The cross-terms are negligible across the whole surface because the coupled surface/body wave term decays algebraically faster than the surface wave term on the cylindrical surface and exponentially faster than the body wave term on the hemispherical surface.     Hence, we may split the  flux  of  radiated waves into a contribution of surface and body waves, respectively.
   
   The energy transported per unit time by body waves through the hemispherical cap $H$ is given by:
   \begin{equation}
   \begin{split}
    \mathcal{E}_b(z_0) =  & \dfrac{\rho \omega^2 c}{2} \int_H |G_b(\mathbf{r},z,z_0)|^2  R^2 d\hat{R}  \\
           = & \dfrac{\rho \omega^2 c}{32 \pi^2 } \int_{2 \pi}  |\psi_b(\mathbf{\hat{R}},z_0)|^2 d\hat{R} ,  \label{ebg}
    \end{split}
   \end{equation}
with $G_b$ the body wave part of Green's function. In the second line of Eq. (\ref{ebg}), the integral is over the space directions subtended by the hemispherical cap. (N.B.: strictly speaking, the total solid angle is not equal to $2 \pi$ because one should remove the directions corresponding to the cylinder. But as noted before, the measure of this set of directions goes to zero as $R_c$ goes to infinity.) The function defined in Eq. (\ref{ebg}) oscillates with depth around the following mean value:
   \begin{equation}
   \langle \mathcal{E}_b \rangle = \dfrac{\rho \omega^2 c}{8 \pi}   \label{meaneb}
   \end{equation}
   Here the brackets may have at least 2 different meanings. The most obvious is an average over depth, as was done in the calculation of the group velocity. But we may also assume that the surface ``scrambles" the phase of the reflected wave $\phi_r$ so that it becomes a random variable. In this scenario, the brackets would mean an average over all realizations of the random reflection process. Upon averaging over phase or depth, the interference pattern between the incident and reflected wave is smoothed out, so that the two approaches yield the same result. Note that the randomization of the phase does not affect energy conservation because the incident flux is still totally reflected. In particular, the discussion following the interpretation of Eq. (\ref{ve}) would still be valid. In practice, the assumption that the phase of the reflected wave is randomized by the surface may not be as unrealistic as it seems.  Observations of reflected $SH$ waves by \citet{kinoshita1993} at borehole stations in Japan indeed suggest that the reflected field is a distorted version of the incident one. This concurs with the general view that the subsurface of the Earth is highly heterogeneous at scales that can be much smaller than the wavelength and brings support to  the idea that  aberrating  fine layers could indeed randomize the phase of the reflected wave as we hypothesize. 
    In what follows, the scrambled-phase assumption will be adopted to simplify the treatment of the reflection of body waves at the surface. 
    
   The energy transported per unit time by surface waves through the lateral cylindrical surface $C$ is given by:
   \begin{equation}
   \begin{split}
    \mathcal{E}_s(z_0) =  & \dfrac{\rho \omega^2 v_g}{2} \int_C |G_s(\mathbf{r},z,z_0)|^2  r d\phi dz  \\
           = & \dfrac{\rho \omega^2 v_g  \alpha^2 e^{-2\alpha z_0}}{4\pi k_s } \int_0^{2 \pi} d\phi \int_0^{h} e^{-2\alpha z} dz  \\
            = & \dfrac{ \rho \omega c^2 \alpha}{4} e^{-2\alpha z_0}
    \end{split}
   \end{equation}
   with $G_s$ the surface wave part of Green's function.  Because $h$ is large compared to $1/\alpha$ the integral over depth may be performed from $0$ to $+\infty$ (the error incurred is exponentially small).
   We find the depth dependent surface-to-body energy ratio:
    \begin{equation}
     R(z_0) = \dfrac{\mathcal{E}_s(z_0)}{\langle \mathcal{E}_b \rangle}  = \dfrac{2 \pi c \alpha }{ \omega} e^{-2\alpha z_0}   \label{eseb}   
    \end{equation}
    Formula (\ref{eseb}) can also be understood in the light of the local density of states $n_{s,b}$ defined as \citep{sheng2006}:
    \begin{equation}
      n_{s,b}(z_0) = - \left.\dfrac{\text{Im} G_{s,b}(\mathbf{r},z_0,z_0)}{\pi} \right|_{\mathbf{r}=\mathbf{0},z=z_0} \times \dfrac{ d k_{s,b}^2 (\omega)}{d\omega},
    \end{equation}
    where $k_{s,b}(\omega)$ stands for the wavenumber of surface or body waves.
    Using the spectral representation (\ref{gex}), one obtains the (exact) formulas:
    \begin{align}
     n_b(z_0) = & \dfrac{\omega^2 }{8 \pi^2 c^3} \int_{2 \pi} |\psi_b(\mathbf{\hat{R}},z_0)|^2 d^2\hat{R}    \label{dosa}  \\
      \langle n_b \rangle = & \dfrac{\omega^2 } {2 \pi c^3}         \label{dosb} \\
     n_s(z_0) = &    \dfrac{\alpha \omega}{c^2} e^{-2\alpha z_0}    \label{doss}
    \end{align}
   which show that  the partitioning of the energy radiated into surface and body waves  by the source $R(z_0)$ is given by the ratio of their local density of states $  n_s(z_0)/\langle n_b \rangle$. 
   Finally, we may compute the partitioning ratio $\mathcal{R}$ between the modal density of surface  and body waves by integrating Eq.(\ref{doss}) over $z$ and taking the ratio with (\ref{dosb}). This yields the simple result:
   \begin{equation} 
    \mathcal{R} =   \int_0^{\infty} \dfrac{ n_s(z)}{ \langle n_b \rangle} dz= \dfrac{\pi c}{\omega} \label{totdos}
   \end{equation}
   where it is to be noted that the modal density ratio $\mathcal{R} $ is independent of the scale length $\alpha$ appearing in the mixed boundary condition of the Helmholtz equation. This result could have been deduced directly from the dispersion relations of body and surface waves using classical mode counting arguments \citep{kittel1976}. It is worth noting that the local density of states (\ref{dosb}) is exactly the same as in the case of the Helmholtz equation in full 3-D space. Although the integral in (\ref{dosa}) is carried over one hemisphere only, each eigenmode $\psi_b$ is composed of an incident and a reflected wave, which -on average- doubles its contribution compared to a single plane wave state.    In the next section, we use our knowledge of the Green's function to derive the scattering properties of surface and body waves including the coupling between the two modes of propagation. 
  \section{Single scattering by a point scatterer}
  In this section, we calculate the energy radiated by a single scatterer in a half-space geometry for incident surface or body waves. For simplicity, we restrict our investigations to point scatterers and employ Born's approximation. The resulting expressions are simplified following the scrambled phase approximation and interpreted in terms of scattering cross-sections. 
  \subsection{Scattering of a surface wave}
  We now consider the following perturbed Helmholtz Eq.:
  \begin{equation}
   \Delta u (\mathbf{r},z)+ k^2(1 + \epsilon a^3 \delta( \mathbf{r})\delta(z-z_s) ) u (\mathbf{r},z)= 0    \label{helmholtzp}
   \end{equation}
   Here $a$ represents the typical linear dimension of the scatterer located at $(0,0,z_s)$ and it is understood that $ka \ll 1$. $\epsilon$ is the local perturbation of inverse squared velocity.  Following the standard procedure \citep{snieder1986}, we look for solutions of Eq. (\ref{helmholtzp}) of the form: $u = u_0 + u_s$, where $u_0 (\mathbf{r},z)= e^{-\alpha z +i \sqrt{\alpha^2 + \omega^2/c^2} x}$ is a surface wave eigenmode of the Helmholtz equation  (the incident field) and $u_s$ is the scattered field. Using the Born approximation, one obtains:
   \begin{equation}
   u(\mathbf{r},z) = u_0 (\mathbf{r},z) - k^2 \epsilon a^3  G(\mathbf{r},z,z_s) u_0(\mathbf{0},z_s)
   \end{equation}
  Introducing the coupling strength $S = k^2 a^3 \epsilon e^{-\alpha z_s}$, one may express the energy radiated by the body waves through the hemispherical cap $H$ (per unit time) as:
 \begin{equation}
   U^{s\rightarrow b} =  \dfrac{\rho \omega^2 S^2 c}{2(4 \pi)^2 R^2} \int_{2\pi} |\psi_b(\mathbf{\hat{R}},z_s)|^2 R^2 d\hat{R}.  \label{eb}
 \end{equation}
  Note that the coupling term $S$ depends on both intrinsic properties of the scatterer -size and strength of perturbations- as well as on the properties of the incident wave -depth dependence of eigenfunction and frequency-.
 
 For a unit amplitude surface wave, the vertically-integrated time-averaged energy flux density is given by 
 \begin{equation}
  | \mathbf{J}_s |= \dfrac{\rho \omega^2 v_g}{4 \alpha}. \label{js}
   \end{equation}
   The ratio of (\ref{eb}) and (\ref{js})  gives the  surface-to-body  scattering cross-section:
   \begin{equation}
    \sigma^{s \rightarrow b}(z_s) =   \dfrac{\alpha c k^4 a^6 \epsilon^2 e^{-2 \alpha z_s}}{ 8 \pi^2 v_g} \int_{2\pi} |\psi_b(\mathbf{\hat{R}},z_s)|^2  d\hat{R}  \label{sigmasb}
       \end{equation}
   This cross-section has unit of length. In the case where the surface scrambles the phase of the reflected wave, we may  compute the mean conversion scattering cross-section by taking the average over the random phase $\phi_r$.
    \begin{equation}
     \langle \sigma^{s \rightarrow b}(z_s) \rangle =  \dfrac{c \alpha k^4 a^6 \epsilon^2 e^{-2\alpha z_s}}{2 \pi v_g} 
   \end{equation}
   Let us remark again that the averaging smoothes out the interference pattern of the body wave eigenfunction $\psi_b$ but does not affect the conservation of energy.
    Furthermore, the averaging procedure makes the scattering pattern  isotropic
   since $\langle  |\psi_b(\mathbf{\hat{R}},z_s)|^2 \rangle =2$ with an equal contribution of upgoing and downgoing waves. 
   The computation of the surface-to-surface scattering cross-section proceeds in a similar way. 
   The surface-wave energy radiated through the lateral surface is given by:
    \begin{equation}
    \begin{split}
       U^{s\rightarrow s} =   &  \dfrac{\rho \omega c^2 \alpha^2 S^2 }{4 \pi } \int_{2\pi} \int^{+ \infty}_0 e^{-2\alpha (z+z_s)} d\phi dz  \\
                                        =   &    \dfrac{\rho \omega c^2 \alpha S^2 e^{-2 \alpha z_s}}{4  }              \label{es}
       \end{split}
 \end{equation}
   Normalizing the result (\ref{es}) by the incident flux yields the surface-to-surface scattering  cross section:
   \begin{equation}
   \sigma^{s\rightarrow s}(z_s) = \dfrac{c \alpha^2 k^3 a^6 \epsilon^2 e^{-4 \alpha z_s} }{v_g},
   \end{equation}
   again with unit of length. 
   If we have a collection of point scatterers with volume density $n$, we may define a surface wave scattering mean free path using the independent scattering approximation as \citep{lagendijk1996,tregoures2002,maeda2008}:
   \begin{equation}
    \dfrac{1}{l^{s}} = \int_0^{\infty} n \left( \sigma^{s \rightarrow s}(z) + \sigma^{s \rightarrow b}(z) \right) dz \label{ils}
   \end{equation}
   The approximate formula (\ref{ils}) neglects all recurrent interactions between the scatterers, which is valid for sufficiently low concentrations of inclusions.
   \subsection{Scattering of body waves}
In the case of an incident body wave mode (\ref{bwm}), the computation of the energy radiated in the form of body or surface waves can be performed as in the previous section. The definition of the scattering cross-section, however, is problematic if we stick to the modal description. Indeed, the vertically integrated energy flux density of a body wave mode, as defined in Eq. (\ref{bwm}), diverges. We must therefore come back to a conventional plane wave description. We first consider the situation where the scatterer is located at a large depth in the half-space.  In this case, it appears reasonable to think that  the scattering of a body wave mode   should be equivalent to the scattering of a plane wave in the  full-space case, at least in a sense to be specified below. To verify the correctness of this assertion, we start by computing the body wave cross-section in absence of a boundary. Assuming a unit amplitude incident plane wave and using again the Born approximation, one may express the scattered energy as:
\begin{equation}
  U_{\text{full}} = \dfrac{\rho \omega^2 c (k^2 \epsilon a^3)^2 }{8 \pi}   , \label{ufull}
\end{equation}
Normalizing the result by the energy flux density of the incident wave:
\begin{equation}
  |\mathbf{J}_b| = \dfrac{\rho \omega^2 c}{2}
\end{equation}
we obtain:
\begin{equation}
 \sigma_{\text{full}} = \dfrac{(k^2 \epsilon a^3)^2}{4 \pi} 
\end{equation}

In the half-space geometry, the incident wave has the form (\ref{bwm}). The total energy scattered in the form of body waves is still given by Eq. (\ref{eb}) provided one redefines the coupling constant as $S = k^2 a^3 \epsilon |\psi_b(\mathbf{\hat{R}}_i,z_s)|^2$, where $\mathbf{\hat{R}}_i$ refers to the incidence direction of the body wave. Eq. (\ref{eb})  differs from formula (\ref{ufull}) as a consequence of the interference between the incident and reflected waves. In the case of a scattering medium, we may expect these interferences to be blurred due to the randomization of the phase by the scattering events. In this scenario, the phase of the incident and reflected waves may be expected to be uncorrelated.  
Upon averaging the result (\ref{eb}) over the random phase of the reflected wave, one obtains:
\begin{equation}
 U^{b \rightarrow b}=   \dfrac{\rho \omega^2 c (k^2 \epsilon a^3)^2 }{4 \pi},
\end{equation}
which is exactly the double of the full-space result (\ref{ufull}).
Keeping in mind that the energy flux of the incident plane wave interacts  twice with the scatterer (direct interaction + interaction after reflexion) and using the assumption that the energy fluxes of incident and reflected waves do not interfere  and may therefore be added, we obtain the result:
\begin{equation}
 \langle \sigma^{b \rightarrow b} \rangle = \sigma_{\text{full}}  \label{sbb}
\end{equation}
 As announced, the average scattering cross-section of body waves in the half space is the same as in the full space for a scatterer located far away from the boundary. Conceptually,  the ``scrambled phase" approximation allows us to extend the result (\ref{sbb}) to a scatterer located at an arbitrary depth in the medium thanks again to the assumption that the incident and reflected waves are statistically independent.  In this scenario, on average, the surface does not modify anything to the scattering of body waves into body waves as compared to the full-space case. 
 
 Following the same approach and approximation, we can calculate the  body-to-surface scattering cross-section. The energy radiated in the form of surface waves is given by:
 \begin{equation}
 U^{b \rightarrow s}=   \dfrac{\rho \omega c^2 \alpha (k^2 \epsilon a^3)^2  e^{-2\alpha z_s} |\psi_b(\mathbf{\hat{R}}_i,z_s)|^2 }{4}
\end{equation}
 After averaging and normalization by the total energy flux (incident + reflected), one finds:
   \begin{equation}
 \langle \sigma^{b \rightarrow s} \rangle(z) =   \dfrac{ \alpha (k^2 \epsilon a^3)^2 e^{-2\alpha z_s}}{2 k} \label{sbs}
 \end{equation}
 Note that all  the remarks pertaining to the mean scattering pattern made after Eq. (\ref{sigmasb}) also apply to the derivation of Eq. (\ref{sbb}) and (\ref{sbs}).
 The scattering cross-sections $\sigma^{b \rightarrow b}$  and $\sigma^{b \rightarrow s}$ have unit of surface. Using the independent scattering approximation again, we conclude that the inverse scattering mean free path of body waves 
 defined as:
  \begin{equation} 
  \dfrac{1}{l^b(z)}=n (\sigma^{b \rightarrow s}(z) + \sigma^{b \rightarrow b}) \label{ilb}
  \end{equation}
  depends on the depth in the medium, as a consequence of the coupling with surface waves. A simple but fundamental reciprocity relation may be established  between the surface-to-body and body-to-surface scattering mean free time.  The later may be expressed as:
 \begin{equation}
  \tau^{b \rightarrow s}(z) = \dfrac{2 k e^{2 \alpha z}}{n c \alpha (k^2 a^3 \epsilon)^2 }
 \end{equation}
 while the former may be obtained after performing the integral over depth in Eq. (\ref{ils}):
 \begin{equation}
    \tau^{s \rightarrow b} = \dfrac{4 \pi}{ n c (k^2 a^3 \epsilon)^2}
 \end{equation}
 The ratio between the two quantities is given by:
 \begin{equation}
  \dfrac{\tau^{s \rightarrow b}}{ \tau^{b \rightarrow s} (z)} = \dfrac{2 \pi  \alpha e^{-2\alpha z}}{k} = R(z)  \label{taubssb}
 \end{equation}
where Eq. ($\ref{eseb}$) has been used.  Eq. (\ref{taubssb}) establishes a link between the surface-to-body versus body-to-surface conversion rates and the local density of states. In the case of elastic waves, a similar relation applies \citep{weaver1990,ryzhik1996}: the ratio between the $P$-to-$S$ and $S$-to-$P$ scattering mean free times is given by the ratio of the density of states of $P$ and $S$ waves. As shown in the next section, the relation (\ref{taubssb}) plays a key role in the establishment of an equipartion between surface and body waves.
 \section{Equation of radiative transfer}
 In this section, we employ standard energy balance arguments to derive a set of coupled equations of RT for surface and body waves  in a half-space containing a uniform distribution of point-scatterers. A notable feature of our formulation is the appearance of the penetration depth of surface waves as a parameter in the Equations.  
 \subsection{Phenomenological derivation}
 Before we establish the transport equation, a few remarks are in order.
It is clear that the transport of surface wave energy is naturally described by a specific surface energy density $\epsilon_s(t,\mathbf{r},\mathbf{\hat{n}})$ where $\mathbf{\hat{n}}$ is a unit vector in the horizontal plane. The surface energy density of surface waves  may in turn be defined as:
\begin{equation}
 \varepsilon_s(t,\mathbf{r}) = \int_{2\pi}\epsilon_s(t,\mathbf{r},\mathbf{\hat{n}})d\hat{n} 
 \end{equation}
  where the integral is carried over all propagation directions in the horizontal plane.  Note that the $\epsilon_s$ symbol should not be confused with the strength of fluctuations in Eq. (\ref{helmholtzp}).
  In contrast with surface waves, the transport of body waves is described by a specific volumetric energy density $e_b(t,\mathbf{r},\mathbf{\hat{k}},z)$ where $\mathbf{\hat{k}}$ is a vector on the unit sphere in 3-D.  The volumetric energy density of body waves is again obtained by integration of the specific energy density over all propagation directions in 3-D:
  \begin{equation}
 E_b(t,\mathbf{r},z) = \int_{4\pi}e_b(t,\mathbf{r},z,\mathbf{\hat{k}})d\hat{k} \label{Eb}
 \end{equation}
 In the usual formulation of transport equations, energy densities have the same unit (either surfacic or volumetric). In order to treat on the same footing surface and body waves, we introduce the following volumetric energy density of surface waves as:
 \begin{equation}
 e_s(t,\mathbf{r},z,\mathbf{\hat{n}})  =  2 \alpha \epsilon_s(t,\mathbf{r},\mathbf{\hat{n}})  e^{-2 \alpha z}  \label{edecay}
 \end{equation}
  It is clear that upon integration of $e_s$  over depth, one recovers the surface density $\epsilon_s$. The exponential decay of the surface wave energy density is directly inherited from the modal shape and implies that the coupling between surface and body waves mostly occurs within a skin layer of typical thickness $1/2 \alpha$. For future reference, we introduce the following notation: 
\begin{equation}  
E_s(t,\mathbf{r},z) = \int_{2 \pi}e_s(t,\mathbf{r},z,\mathbf{\hat{n}}) d\hat{n} 
\label{Es}
\end{equation}
  to represent the volumetric energy density of surface waves, consistent with Eq. (\ref{Eb}).
   Note that the total energy density at a given point will be defined as the sum $(E_b(t,\mathbf{r},z) +E_s(t,\mathbf{r},z) )$, thereby implying the incoherence between the two types of waves.
  
With this definition of the surface wave energy density,  the phenomenological derivation of the radiative transport equation  follows exactly the same procedure as in the multi-modal case \citep{turner1994}. A beam of energy followed along its path around direction $\mathbf{\hat{n}}^i$ is affected by (1)  conversion of energy into other propagating modes and/or  deflection into other propagation directions $\mathbf{\hat{n}}^o \neq \mathbf{\hat{n}}^i$; (2) a gain of energy thanks to the reciprocal process: a wave with mode $i$ propagating in direction $\mathbf{\hat{n}}^i$ can be converted into a wave with mode $o$ propagating in direction $\mathbf{\hat{n}}^o$ by scattering. 
In the general case of finite size scatterers we may anticipate scattering to be  anisotropic.  To describe such an  angular dependence of the scattering process, we may introduce normalized phase functions $p^{i \rightarrow o}(\mathbf{\hat{n}}^o,\mathbf{\hat{n}}^i)$. 
 The phase function   may be understood as the probability that a wave of mode $i$ propagating in direction $\mathbf{\hat{n}}^i$ be converted into a wave of mode $o$ propagating in direction  $\mathbf{\hat{n}}^o$.  To be interpretable probabilistically, its integral over all outgoing directions ($\mathbf{\hat{n}}^o$) should equal 1. Note that in addition to the incoming and outgoing propagation directions, the phase function may also depend on the depth in the medium  as a consequence of the presence of the surface. In the case of point scatterers,  this complexity disappears thanks to  the scrambled phase assumption and will therefore not be considered in our formalism.   

A detailed local balance of energy  yields the following system of coupled transport equations:
 \begin{equation}
 \begin{split}
\left( \partial_t +  v_g  \mathbf{\hat{n}}\cdot \nabla  \right)   e_s(t,\mathbf{r},z,\mathbf{\hat{n}}) =  & - \dfrac{e_s(t,\mathbf{r},z,\mathbf{\hat{n}})}{\tau^s} 
+\dfrac{1}{\tau^{s \rightarrow s}} \int_{2 \pi} p^{s \rightarrow s}(\mathbf{\hat{n}},\mathbf{\hat{n}}')  e_s(t,\mathbf{r},z,\mathbf{\hat{n}}') d\hat{n}'   \\
&+ \dfrac{1}{\tau^{b \rightarrow s}(z)} \int_{4 \pi} p^{b \rightarrow s}(\mathbf{\hat{n}},\mathbf{\hat{k}}';z)  e_b(t,\mathbf{r},z,\mathbf{\hat{k}}') d\hat{k}' 
 + s_s(t,\mathbf{r},z,\mathbf{\hat{n}}) \\ 
\left( \partial_t +  c \mathbf{\hat{k}}\cdot \nabla   \right)   e_b(t,\mathbf{r},z,\mathbf{\hat{k}}) =&  - \dfrac{e_b(t,\mathbf{r},z,\mathbf{\hat{k}})}{\tau^b(z)} 
+ \dfrac{1}{\tau^{b \rightarrow b}} \int_{4 \pi} p^{b \rightarrow b}(\mathbf{\hat{k}},\mathbf{\hat{k}}')  e_b(t,\mathbf{r},z,\mathbf{\hat{k}}') d\hat{k}'   \\
&+ \dfrac{1}{\tau^{s \rightarrow b}} \int_{2 \pi} p^{s \rightarrow b}(\mathbf{\hat{k}},\mathbf{\hat{n}}')  e_s(t,\mathbf{r},z,\mathbf{\hat{n}}') d\hat{n}'  
  + s_b(t,\mathbf{r},z,\mathbf{\hat{n}})  \label{rtb}
  \end{split}
  \end{equation}
  where the terms $s_{s,b}$ represent sources of surface and body waves.  To take into account the reflection of body waves at the surface,  the sytem (\ref{rtb}) is  supplemented with the following boundary condition for the energy density $e_b$:
 \begin{equation}
    e_b(t,\mathbf{r},0,\mathbf{\hat{k}}) = e_b(t,\mathbf{r},0,\mathbf{\hat{k}}_r)     \label{bc}
 \end{equation}
 where $\mathbf{\hat{k}}_r$ is the mirror image of the incident direction $\mathbf{\hat{k}}$  $(\mathbf{\hat{k}}\cdot \mathbf{\hat{z}} <0)$ across the horizontal plane $z=0$.  The boundary condition (\ref{bc}) is compatible with the assumption that the incident and reflected waves are statistically independent.

 In the simple case of a unit point-like  source at depth $z_0$, the  terms $s_s$ and $s_b$  are given respectively by:
  \begin{equation}
   \begin{split}
    s_s(t,\mathbf{r},z,\mathbf{\hat{n}}) = & \dfrac{2 \alpha R(z_0)  e^{-2 \alpha z}  \delta(\mathbf{r)}}{2 \pi(1 + R(z_0))}   \\
    s_b(t,\mathbf{r},z,\mathbf{\hat{k}}) = & \dfrac{ \delta(z -z_0) \delta(\mathbf{r)}}{4 \pi(1 + R(z_0))} 
    \end{split}   \label{sssb}
  \end{equation}
   where $R(z_0)$ is the energy partitioning ratio defined in  Eq. (\ref{eseb}). 
   Note that $s_s$ follows the same exponential decay as $e_s$ with depth $z$ (see Eq. \ref{edecay}), which takes into account the vertical dependence of the surface wave eigenfunction. In Eq. (\ref{sssb}), the complex dependence of the body wave radiation with depth has been simplified by averaging the exact result (\ref{ebg}) over the random phase of the reflected wave.  As a consequence, the energy radiation at the source covers uniformly the whole sphere of space directions in 3-D. 
   
   A similar remark applies to  the scattering from body waves to body waves, which have been treated as if the surface was absent in Eq. (\ref{rtb}).  Again, this approximation is admissible if the scattered  upgoing and downgoing energy fluxes are statistically independent, a condition which is guaranteed by the randomization of the phase of the waves upon reflection at the surface in our model.
As discussed in the previous section,    in the case of point scatterers the phase averaging procedure makes  all scattering processes isotropic which allows us to simplify the system of Eqs (\ref{rtb})  by evaluating the scattering integrals on the right-hand side of Eq. (\ref{rtb}):
  \begin{equation}
  \begin{aligned}
  \left( \partial_t +  v_g  \mathbf{\hat{n}}\cdot \nabla  \right)   e_s(t,\mathbf{r},z,\mathbf{\hat{n}}) =  & - \dfrac{e_s(t,\mathbf{r},z,\mathbf{\hat{n}})}{\tau^s}  + \dfrac{E_s(t,\mathbf{r},z)}{2 \pi \tau^{s \rightarrow s}}  +    \dfrac{E_b(t,\mathbf{r},z)}{2 \pi \tau^{b \rightarrow s}(z)}  + s_s(t,\mathbf{r},z,\mathbf{\hat{n}}) \\
    \left( \partial_t +  c  \mathbf{\hat{k}}\cdot \nabla  \right)   e_b(t,\mathbf{r},z,\mathbf{\hat{n}}) =  & - \dfrac{e_b(t,\mathbf{r},z,\mathbf{\hat{k}})}{\tau^b(z)}  + \dfrac{E_b(t,\mathbf{r},z)}{4\pi \tau^{b \rightarrow b}}  +    \dfrac{E_s(t,\mathbf{r},z)}{4 \pi \tau^{s \rightarrow b}}  + s_b(t,\mathbf{r},z,\mathbf{\hat{k}}) 
  \end{aligned} \label{rtiso}
  \end{equation}
   where we have used the normalization condition of the phase functions. 
   The decreasing efficacy of scattering conversions with depth is guaranteed by the exponential decay of $\tau^{b \rightarrow s}(z)^{-1}$ and $E_s(t,\mathbf{r},z)$ in the first and second Eq. of the coupled sytem (\ref{rtiso}), respectively. It is worth recalling that  the depth dependence of the  scattering mean free times  of body waves $\tau^{b}(z)$ and $\tau^{b \rightarrow s}(z)$ is caused by the coupling with surface waves and not by a stratification of heterogeneity.
  \subsection{Energy conservation and equipartition}
 A self-consistent formulation of coupled transport equations should verify two elementary principles: energy conservation and equipartition.
 To demonstrate these properties from the basic set of equations (\ref{rtb}) or (\ref{rtiso}), we proceed in the usual fashion  \citep{turner1994}. An integration of
 each equation over all possible propagation directions $\mathbf{\hat{n}}$ (in 2-D) or $\mathbf{\hat{k}}$ (in 3-D) yields:
 \begin{equation}
 \begin{aligned}
 \partial_t    E_s(t,\mathbf{r},z)   + \nabla \cdot \mathbf{J}_s(t,\mathbf{r},z) =  &  - \dfrac{ E_s(t,\mathbf{r},z)}{\tau^{s\rightarrow b}} 
  + \dfrac{E_b(t,\mathbf{r},z)}{\tau^{b \rightarrow s}(z)}   \\ 
\partial_t    E_b(t,\mathbf{r},z)   + \nabla \cdot \mathbf{J}_b(t,\mathbf{r},z) = &    - \dfrac{E_b(t,\mathbf{r},z)}{\tau^{b \rightarrow s}(z)}  
  + \dfrac{E_s(t,\mathbf{r},z)}{\tau^{s \rightarrow b}}   \label{Econsb} ,
  \end{aligned},
\end{equation}
 where we have used the definition of the mean free times on the RHS of (\ref{Econsb}).
 Note that we have dropped the source terms since they are not essential to our argumentation.
 In Eq. (\ref{Econsb}), we have introduced the energy  flux density vector of surface and body waves:
 \begin{equation}
 \begin{aligned} 
 \mathbf{J}_{s} &= \int_{2 \pi}  e_s(t,\mathbf{r }, z, \mathbf{\hat{n}}) v_g \mathbf{\hat{n}} d\hat{n} \\
 \mathbf{J}_{b} &= \int_{4 \pi}  e_b(t,\mathbf{r }, z, \mathbf{\hat{k}}) c \mathbf{\hat{k}} d\hat{k}
 \end{aligned}
 \end{equation}
  Note that  $\mathbf{J}_{s}$ is contained in a horizontal plane. Upon integration over the whole half-space, the terms which contain the current density vectors can be converted into surface integrals that  vanish. Denoting by a double bar an integration over $\mathbf{r}$ and $z$, and summing the two Eqs of the system  (\ref{Econsb}) leaves us with:
 \begin{equation}
   \partial_t   \left( \bar{\bar{E}}_s (t) + \bar{\bar{E}}_b (t) \right) =0 \label{econstot}
   \end{equation}
   which demonstrates the conservation of energy.
   
   In order to prove the existence of equipartition, we integrate each Eq. of the system (\ref{Econsb}) over the horizontal plane and from the surface to a finite depth $h$, typically large compared to the surface wave penetration depth. To simplify the derivation, we assume that at $z=h$ lies a perfectly reflecting surface through which no energy can flow. Because $\alpha h \gg  1$, the medium may be still be considered as a half-space in the treatment of surface waves. This leads us to:
  \begin{equation}
 \begin{aligned}
 \partial_t   \bar{\bar{E}}_s(t)  =   & - \dfrac{  \bar{\bar{E}}_s(t)}{\tau^{s\rightarrow b}} 
  + \int_0^h \dfrac{\bar{E}_b(t,z)}{\tau^{b \rightarrow s}(z)}dz     \\
\partial_t   \int_0^h \bar{E}_b(t,z) dz  =   & - \int_0^h \dfrac{\bar{E}_b(t,z)}{\tau^{b \rightarrow s}(z)}  dz
  + \dfrac{\bar{\bar{E}}_s(t)}{\tau^{s \rightarrow b}}   \label{Econsb2} ,
\end{aligned}
\end{equation}
 where the single bar denotes an integration over $\mathbf{r}$ only. Note that $ \bar{\bar{E}}_s$ is the total  energy of surface waves (up to an exponentially small correction term), in contrast with $\bar{E}_b$ which represents the body wave energy per unit depth. Our goal is not to solve the system of Eq. (\ref{Econsb2}) in its full generality but rather to exhibit an equipartition solution. In this regime, we expect the distribution of body wave energy to become independent of depth. Hence we look for solutions of the system (\ref{Econsb2})
 of the form:
 \begin{gather}
        \begin{pmatrix}   
        \bar{\bar{E}}_s(t)  \\
          \bar{E}_b(t,z) 
         \end{pmatrix}      
            =   \begin{pmatrix}
                 \bar{\bar{E}}^0_s   \\
                   \bar{ E}^0_b
                 \end{pmatrix}
                    e^{-\lambda t}       \label{ansatz}
 \end{gather}
 Reporting the ansatz (\ref{ansatz}) into the integro-differential system (\ref{Econsb2}), one arrives at a linear and homogeneous system of algebraic equations. A non-zero solution is obtained only if the determinant of the sytem vanishes which implies:
 \begin{equation}
        \lambda \left(\lambda  - \dfrac{1}{\tau^{s \rightarrow b}}  -\dfrac{1}{h} \int_0^h \dfrac{dz}{\tau^{b \rightarrow s}(z)}   \right) = 0
  \end{equation}      
 One of the solutions is given by $\lambda = 0$ which corresponds to the asymptotic equipartition state such as:
    \begin{equation}
      \dfrac{  \bar{\bar{E}}^0_s  }{ \bar{ E}^0_b} =   \int_0^h \dfrac{\tau^{s \rightarrow b}}{\tau^{b \rightarrow s}(z)} dz \approx \int_0^{+\infty} R(z)dz= \mathcal{R} \label{equip}
    \end{equation}
    where Eq. (\ref{totdos}) and  (\ref{taubssb}) have been used, and the depth integral is again extended to $+\infty$ thanks to the assumption $\alpha h \gg 1$. Eq. (\ref{equip}) illustrates that the ratio between the energy density of surface and body waves approaches the ratio of their density of states $\mathcal{R}$ at long lapse-time. Note that the energy density is not perfectly homogenized spatially, even at equipartition, as a consequence of the decay of the surface wave eigenfunction with depth. This may be related to the depth-dependence of the density of states near the boundary \citep[see e.g.][]{hennino2001}.
    The result (\ref{equip}) could have been predicted using the usual concept of equipartition which states that when filtered around a narrow frequency band, all the propagating modes of a diffuse field should be excited to equal energy \citep{weaver1982}.
    In the case where the medium is unbounded at depth, there will be a  flux of energy across the lower boundary $z=h$ which vanishes as the lapse-time increases. As demonstrated through numerical simulations later in this paper, equipartition also sets in in this configuration, though  probably more slowly than in the slab geometry.    
\subsection{Diffusion Approximation}
Having established the existence of an equipartition state, we now derive a diffusion approximation for the transport process. At the outset, it should be clear that the volumetric energy density $E_s$ may not be the solution of a diffusion equation because it exhibits a decay with depth which is  inherited from the modal shape and therefore independent of the scattering properties. To circumvent the difficulty, we derive a closed diffusion equation for $E_b$ from which we subsequently deduce the energy density of surface waves. To simplify the calculations, we assume that the scattering is isotropic. Proceeding in the usual fashion, we expand the specific energy density into its first two angular moments \citep{akkermans2007}:
\begin{equation}
\begin{aligned}
e_s(t,\mathbf{r},z,\mathbf{\hat{n}}) = & \dfrac{E_s(t,\mathbf{r},z)}{2 \pi} + \dfrac{\mathbf{J}_s(t,\mathbf{r},z)\cdot \mathbf{\hat{n}}}{\pi} \\
e_b(t,\mathbf{r},z,\mathbf{\hat{k}}) = & \dfrac{E_b(t,\mathbf{r},z)}{4 \pi} + \dfrac{3 \mathbf{J}_b(t,\mathbf{r},z)\cdot \mathbf{\hat{k}}}{4 \pi}
\end{aligned}. \label{momentexp}
\end{equation}	
Multiplying, respectively, the first and second line of Eq. (\ref{rtiso}) by the unit vectors $\mathbf{\hat{n}}$ and $\mathbf{\hat{k}}$, integrating over all possible directions and employing the moment expansion (\ref{momentexp}), we obtain the following set of Equations:
\begin{equation}
\begin{aligned}
\dfrac{\partial_t \mathbf{J}_s(t,\mathbf{r},z)}{v_g}  + \dfrac{v_g}{2}\nabla_{\parallel} E_s(t,\mathbf{r},z) = & -\dfrac{\mathbf{J}_s(t,\mathbf{r},z)}{v_g \tau^s} \\ 
\dfrac{\partial_t \mathbf{J}_b(t,\mathbf{r},z)}{c}  + \dfrac{c}{3}\nabla E_b(t,\mathbf{r},z) = & -\dfrac{\mathbf{J}_b(t,\mathbf{r},z)}{c \tau^b}
\end{aligned} ,\label{prefick}
\end{equation} 
where $\nabla_{\parallel}$ denotes the gradient operator in the horizontal plane. 
Note that the expansions (\ref{momentexp}) are used only to evaluate the integral of the gradient on the left-hand side of the RT Equation. All other terms follow directly either from the definition of the  energy flux density vector or the assumption of isotropic scattering. 
Our final approximation consists in neglecting the derivative of the current vector with respect to time which yields the equivalent of Ohm's law for the mutiply-scattered waves:
\begin{equation}
\begin{aligned}
   \mathbf{J}_s(t,\mathbf{r},z)= & - D_s \nabla_{\parallel} E_s(t,\mathbf{r},z) \\
\mathbf{J}_b(t,\mathbf{r},z)= & - D_b(z) \nabla E_b(t,\mathbf{r},z)
\end{aligned} \label{ohm}
\end{equation}  
where the following notations have been introduced:
\begin{equation}
\begin{aligned}
    D_s = & \dfrac{v_g^2 \tau^s}{2}  \\
    D_b(z) = & \dfrac{c^2 \tau^b(z)}{3}
\end{aligned}
\end{equation}
Note that the total reflection condition (\ref{bc}) imposes that there is no net flux of body waves across the surface $z=0$.
To make progress, we now invoke the equipartition principle to fix the ratio between the energy densities of surface and body waves:
\begin{equation}
\dfrac{E_s(t,\mathbf{r},z)}{E_b(t,\mathbf{r},z)} = R(z) \label{equip2}
\end{equation}
in agreement with Eq. (\ref{equip}). This allows us to express the total energy flux as:
\begin{equation}
\mathbf{J} = \mathbf{J}_b + \mathbf{J}_s = -(R(z)D_s + D_b(z)) \nabla_{\parallel} E_b(t,\mathbf{r},z) + D_b(z) \mathbf{\hat{z}} \partial_z  E_b(t,\mathbf{r},z)
 \label{ohm2}
\end{equation}  
which may be interpreted as a generalization of Ohm's law for diffuse waves. Eq (\ref{ohm2}) demonstrates that the coupling between surface and body waves renders the energy transport both depth-dependent and anisotropic. It is worth emphasizing that  depth-dependence and anisotropy are caused neither by specific orientations/shapes of the scatterers nor by the non-homogeneity of the statistical properties. As further discussed below, these properties stem from the coupling between body and surface wave modes. 

The conservation equation for the total energy $E = E_s + E_b$ excited by a point source at $t=0$ is obtained by taking the sum of the set of Eqs (\ref{Econsb}):
\begin{equation}
 \partial_t E(t,\mathbf{r},z) + \nabla \cdot J(t,\mathbf{r},z) = \delta(t) \delta(z-z_0) \delta(\mathbf{r}),
\end{equation}
where the right-hand side now contains the source term with $z_0$ the source depth. Making use of Eqs (\ref{equip2}) and (\ref{ohm}), we obtain the following diffusion-like equation verified by the body wave energy density:
\begin{equation}
\begin{split}
  \partial_t E_b(t,\mathbf{r},z) - \nabla_{\parallel} \cdot \left( \dfrac{D_b(z) +R(z)D_s}{1+R(z)} \nabla_{\parallel} E_b(t,\mathbf{r},z)\right) -\dfrac{1}{1+R(z)}\partial_z \left( D_b(z) \partial_z E_b(t,\mathbf{r},z)\right)  =  \dfrac{\delta(t)\delta(\mathbf{r})\delta(z-z_0)}{1+R(z_0)}
\end{split} \label{diffus1}
\end{equation} 
 The last term on the left-hand side of Eq. (\ref{diffus1}) differs from the traditional form for the diffusion model due to the $(1 + R(z))^{-1}$ factor in front of the derivative operators. Actually, this difference is purely formal as may be shown by the change of variable $z \rightarrow z' $ where $z'$ is defined as:
 \begin{equation}
 \begin{split}
 z' = & \int_0^z (1 + R(x))dx, \\
    = & z +\dfrac{\pi}{k} \left(e^{-2\alpha z} - 1 \right)
 \end{split} \label{newvar}
  \end{equation}
  where Eq. (\ref{eseb}) has been used.
In the new variables, Eq. (\ref{diffus1}) may be rewritten as:
\begin{equation}
   \partial_t E'_b(t,\mathbf{r},z') - \nabla_{\parallel} \cdot \left( D_{\parallel}(z') \nabla_{\parallel} E'_b(t,\mathbf{r},z')\right) -\partial_{z'} \left( D_{\perp}(z') \partial_{z'} E'_b(t,\mathbf{r},z')\right)  =\\\delta(t)\delta(\mathbf{r})\delta(z'-z'_0) \label{diffus2}
\end{equation}
In Eq. (\ref{diffus2}), we have introduced the notations $E'_b(t,\mathbf{r},z') = E_b(t,\mathbf{r},z) $, $z'_0 = \int_0^{z_0} (1 + R(x))dx$, as well as the following definitions of the horizontal and vertical diffusivities:
\begin{equation}
\begin{aligned}
D_{\parallel}(z') = & \dfrac{D_b(z) + R(z)D_s}{1 + R(z)}  \\
D_{\perp}(z')  =  & D_b(z)(1+R(z)) 
\end{aligned} \label{diffusivites}
\end{equation}
  Hence, a simple change of scale in the vertical direction reduces Eq.(\ref{diffus1}) to the conventional diffusion Eq. (\ref{diffus2}). Note that the $(1 +R(z_0))^{-1}$ factor on the right-hand side has been absorbed by the change of variable (\ref{newvar}). 
  
We explore the consequences of Eq. (\ref{diffusivites}) by first considering the case $z' \to \infty$. According to Eq. (\ref{newvar}), this implies $z' \approx z $. Since the partitioning ratio $R(z)$ goes to zero at large depth in the medium, Eq. (\ref{diffusivites}) indicates that the vertical and horizontal diffusivities become equal and the diffusion tensor isotropic. Furthermore, because the coupling between surface and body waves is negligible at large depth, its magnitude tends to the constant value $D_b^{bulk} = c^2 \tau^{b \rightarrow b}/3$, as expected on physical grounds. In other words, the diffusion process at depth is governed by a simple 3-D diffusion equation for body waves with diffusion constant $D_b^{bulk}$. This in turn suggests that at long lapse-time, the coda should decay as $t^{-3/2}$ in a non-absorbing medium. This point will be further substantiated by numerical simulations.
   
In the vicinity of the surface $z = O(\alpha^{-1})$, Eq. (\ref{diffusivites}) shows that the diffusivity of coupled body and surface waves is both depth dependent and non-isotropic. 
The origin of the $z$-dependence is clear since the efficacy of the coupling between surface and body waves decays exponentially with depth. In the vicinity of the surface, the anisotropy stems from the transport of a fraction of the energy by surface waves whose velocity and scattering mean free time differ from the one of body waves. In Eq. (\ref{diffusivites}) the transverse diffusivity is recognized as a weighted average of the surface and body wave diffusivities with coefficients dictated by the equipartition principle. The vertical diffusivity is -up to the $(1+R(z))$ pre-factor inherited from the change of scale in the vertical direction- equal to the diffusivity of body waves. 
In the next section, we illustrate the transport process of coupled body and surface waves by numerically simulating the system of Eq. (\ref{rtb}).
 \section{Monte-Carlo Simulations}

 In this section, we explore some of the key features of our model with the aid of numerical simulations. The approach to equipartition as well as the role of mode coupling in the coda excitation are illustrated. 
 \subsection{Overview of the method}
 As outlined in introduction, Monte-Carlo simulations have been used for more than thirty years in seismology to simulate the transport of seismic energy in heterogeneous media. Our approach to the solution of the coupled set of transport equations (\ref{rtiso}) for surface and body waves follows closely the approach of \citet{margerin2000}, with some appropriate modifications which we outline briefly. 
 
 Energy transport is modeled by the simulation of a large number of random trajectories of particles or seismic phonons \citep{shearer2004}. Each particle is described by its mode, position, propagation direction and time. The initial mode of propagation is randomly selected, following the source energy partitioning ratio (\ref{eseb}), and the initial propagation direction is a uniformly distributed random vector in 2-D (resp. 3-D) for surface (resp. body) waves. 
 Note that when the particle is of surface type, the particle propagates in a horizontal plane and its exact depth is immaterial. In fact, we may say that a particle of surface type is present at all depth with a probability distribution given by $p_s(z)= 2\alpha \exp(-2\alpha z)$ inherited from the modal shape. The lapse-time to the first scattering event is randomly determined and obeys a simple exponential distribution when the particle represents surface waves. In the case of body waves, the selection process is more complicated because their scattering mean free time depends on the depth in the medium. To address this difficulty, we employ the method of delta collisions, which simulates in a simple and exact way a completely general distribution of scattering mean free time. We will not detail the method here and refer the interested reader to the pedagogical treatment by \citet{lux1991}. At each scattering event, the mode of the particle is randomly selected by interpreting probabilistically Eqs (\ref{ils}) and (\ref{ilb}) defining the scattering attenuations. As an example, $(1/l^{s \rightarrow b})/(1/l^s)$ may be interpreted as the transition probability from a surface to a body wave mode. Note that when such an event occurs, the particle is reinjected at a random depth in the medium following the probability distribution $p_s(z)$.  
  To obtain energy envelopes, the position and mode of the particle is monitored on a cylindrical grid at regular time intervals. The local energy density is estimated by averaging the number of particles per cell over a sufficiently large number of random walks. For accuracy, it is important that the cells be relatively small compared to the shortest mean free path in the medium.   
 \subsection{Numerical results}
\begin{figure}
\centerline{\includegraphics[width=0.94\linewidth]{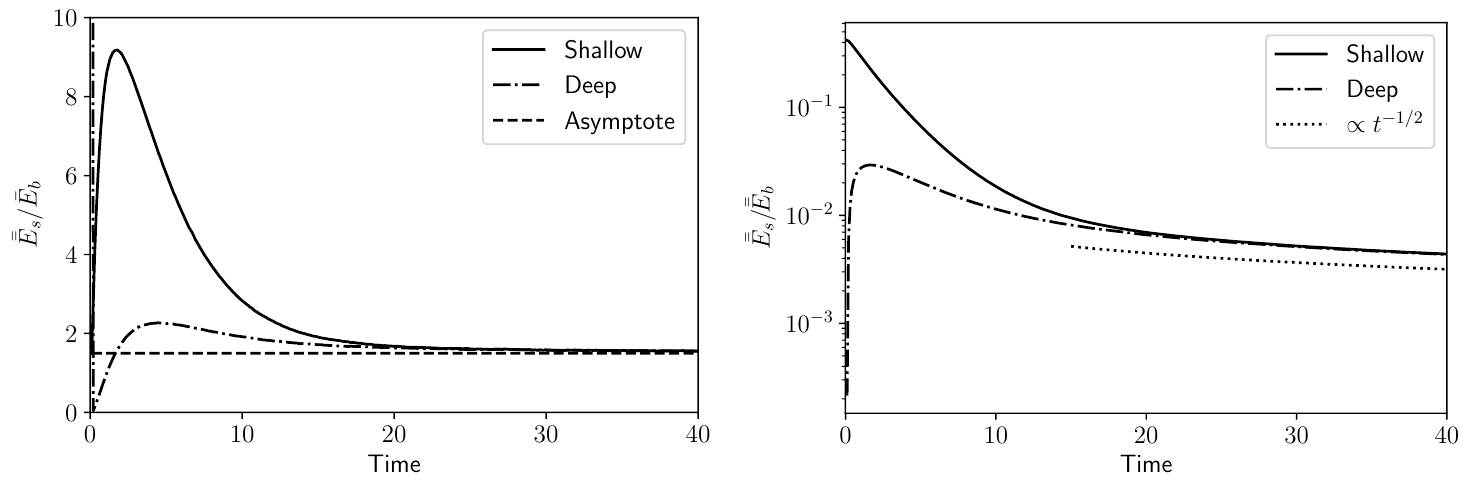}}
 \caption{Local and global evolution of energy partitioning for a shallow (solid line) and a deep (dash-dotted line) source in a heterogeneous half-space filled with point scatterers. The horizontal time scale is the mean free time for surface wave scattering. The penetration depth of surface waves is $\alpha^{-1} =1$ km and the angular frequency is $\omega=2\pi$ Hz. The shallow and deep source are located, respectively, at depth $\alpha^{-1}$ and $5 \alpha^{-1}$. (Left): Temporal evolution of the ratio between the total energy  of surface waves ($\bar{\bar{E}}_s$) and the energy density of body waves integrated over a horizontal plane ($\bar{E}_b$). The body wave energy is evaluated at the surface and averaged over a depth $\Delta z=1$km. The asymptote (left) is the prediction of Eq. (\ref{totdos}). (Right) Temporal evolution of the ratio between the total energy of surface ($\bar{\bar{E}}_s$) and body waves ($\bar{\bar{E}}_b$). The dotted line shows an algebraic $t^{-1/2}$ decay.} \label{figequip}
 \end{figure}

Figure \ref{figequip} illustrates the striking difference between the \emph{global} and \emph{local} partitioning of the seismic energy into surface and body waves. The following parameters have been employed in the simulation: $\alpha=1$km$^{-1}$, $c=3$km/s and $\tau^{s\rightarrow s} = 20$s,  $\tau^{s\rightarrow b} = 30$s,  $\tau^{b\rightarrow b} = 30$s, $\tau^{b\rightarrow s}(z) = 10 \exp(2z)$s. 
Note that in our model the group velocity of surface waves $v_g \approx 3.32$km/s is slightly faster than the speed of propagation of body waves. Two source depths are considered: a relatively shallow one ($z_0 =1$km) and a deep one ($z_0=5$km), which radiate  approximately $29\%$ and $0.01\%$ energy as surface waves, respectively.

 \begin{figure}
 \centerline{\includegraphics[width=0.58\linewidth]{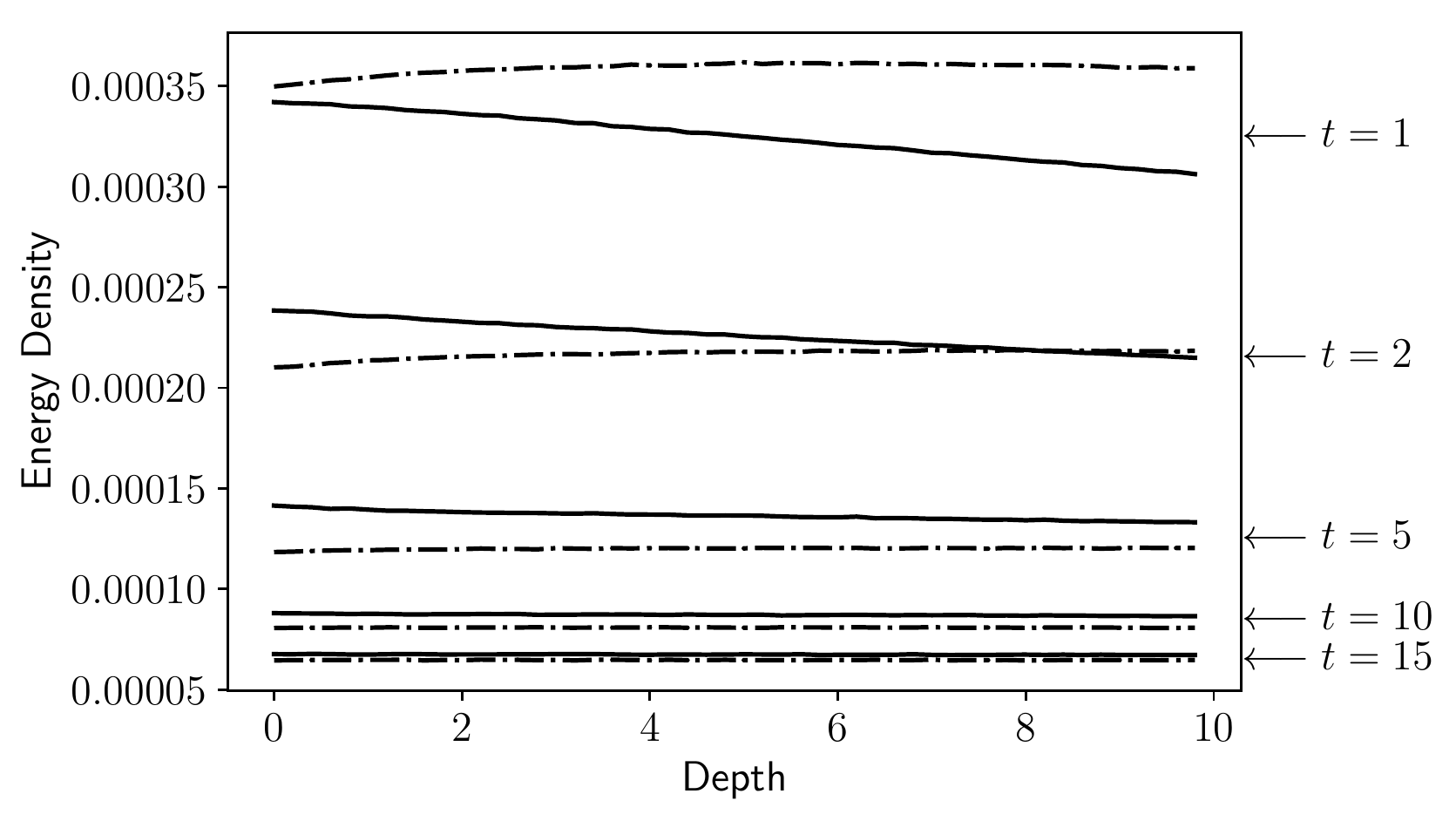}}
\caption{Horizontally-averaged energy density of body waves $\bar{E}_b$ as a function of depth in a heterogeneous half-space filled with point scatterers. The energy is averaged over a depth range $\Delta z=\alpha^{-1}/5=0.2$km and the horizontal axis shows the depth normalized by $\alpha^{-1}$. The solid and dashed  lines correspond to a shallow source ($z_0=1$km) and a deep  source ($z_0=5$km), respectively. The lapse time in the coda in mean free time unit is indicated on the right of the corresponding curves. } \label{fighomog}
\end{figure}

 On the left, we show the temporal evolution of the ratio between the total energy of surface waves  $\bar{\bar{E}}_s$ (see the remarks before Eq. \ref{econstot}  for a reminder of the notations), and the horizontally-integrated energy density of body waves $\bar{E}_b$ at the surface $z=0$, averaged over a depth range $\Delta z=1$km. Hence, the ratio $\bar{\bar{E}}_s/\bar{E}_b$ has unit of inverse length. 
 Independent of the source depth, we find that the partitioning of the energy density at the surface -into surfacic energy of surface waves and volumetric energy of body waves-  converges toward the prediction of equipartition theory, at long lapse-time in the coda (see Eq. \ref{totdos} and \ref{equip}). This numerical result confirms that the analysis of equipartition given in the previous section in slab geometry extends to the half-space geometry. Furthermore,  we find that the surface-to-body energy ratio overshoots the prediction of equipartition theory for the two sources at short lapse-time, by a factor which decreases with the source depth $z_0$. 
 
  The stabilization of the local energy density ratio of surface and body waves at the surface of the half-space is to be contrasted with the evolution of the \emph{global} partitioning of the energy into surface and body wave modes. 
 Figure \ref{figequip} (right) shows that after a few mean free times, most of the energy is carried in the form of body waves in the medium. The Figure also suggests that the global transfer of energy from surface waves to body waves occurs at a rate proportional to $t^{-1/2}$ at long lapse-time. 
 
 Further insight into the equipartition process is offered in Figure \ref{fighomog}, where we show the depth dependence of the horizontally-averaged  body wave energy density at different lapse-time in the coda. All the parameters of the simulation are the same as in Figure \ref{figequip}, except for the much finer spatial resolution $\Delta z= \alpha^{-1}/5=0.2$ km, which allows us to track processes that occur in the skin layer where the coupling between surface and body waves occurs. We observe that after roughly 10 mean free times, the depth distribution of body wave energy becomes homogeneous over a depth range at least as large as $10\alpha^{-1}$, independent of the source depth. This simulation therefore confirms the { theoretical} analysis performed in slab geometry. The  homogenization of the energy of body waves is a dynamic process: the energy density of surface waves increases exponentially near the surface, thereby generating a larger amount of body-wave converted energy; this process is compensated by the exponential increase of the conversion rate from body to surface waves, which eventually yields an equilibrium.  Note that the total energy density does \emph{not} homogenize with depth, due to the exponential decay of the surface wave eigenfunction with depth.

   \begin{figure}
 \centerline{\includegraphics[width=0.98\linewidth]{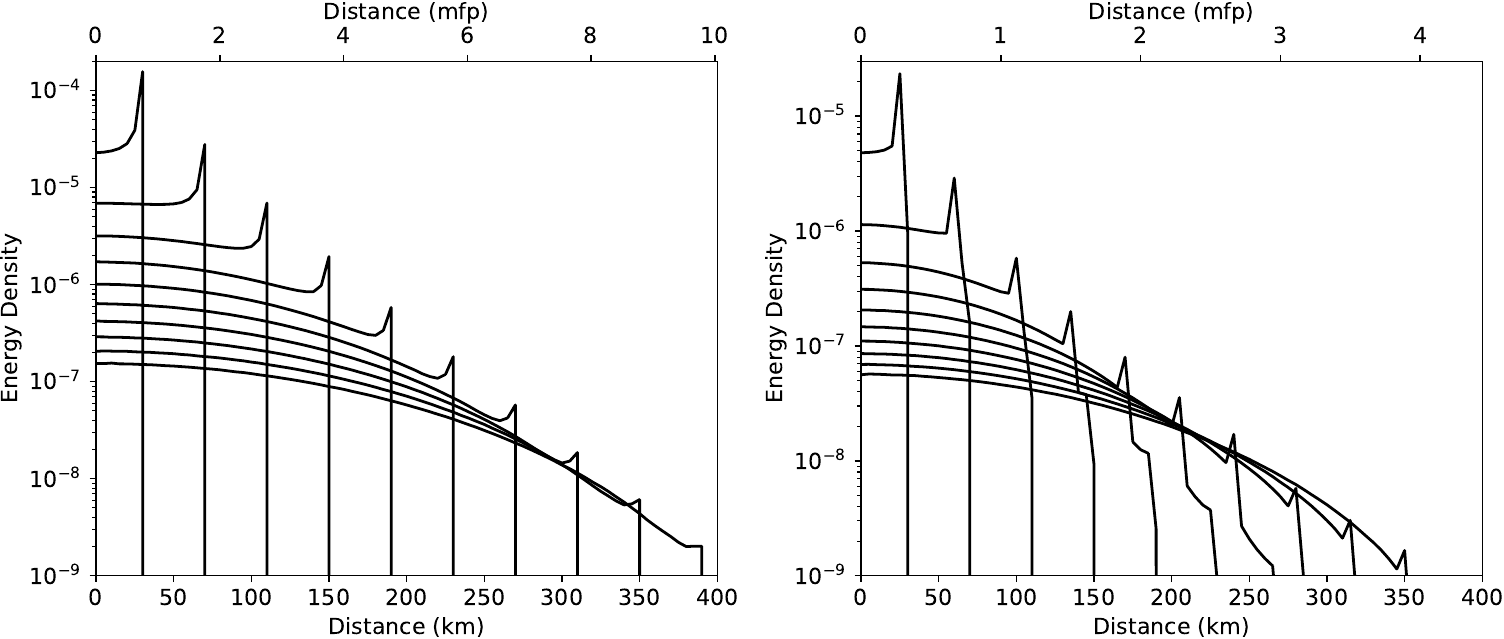}}
 \caption{Snapshots of the energy density of surface waves $\epsilon_s$ (left) and of the volumetric energy density of body waves $E_b$ (right) at the surface  of a heterogeneous half-space filled with point scatterers in the case of a shallow source ($z_0 = \alpha^{-1} $). {The horizontal axes are in units of the scattering mean free path of surface waves (left, top), body waves (right, top) and in kms (bottom). For body waves, we take the mean free path value in the bulk of the medium ($z \to \infty$). } The energy is averaged over cylindrical shells of width $\Delta r = 5$km and deph $\Delta z =5$km. }\label{snapshallow}
 \end{figure}
 \begin{figure}
 \centerline{\includegraphics[width=0.98\linewidth]{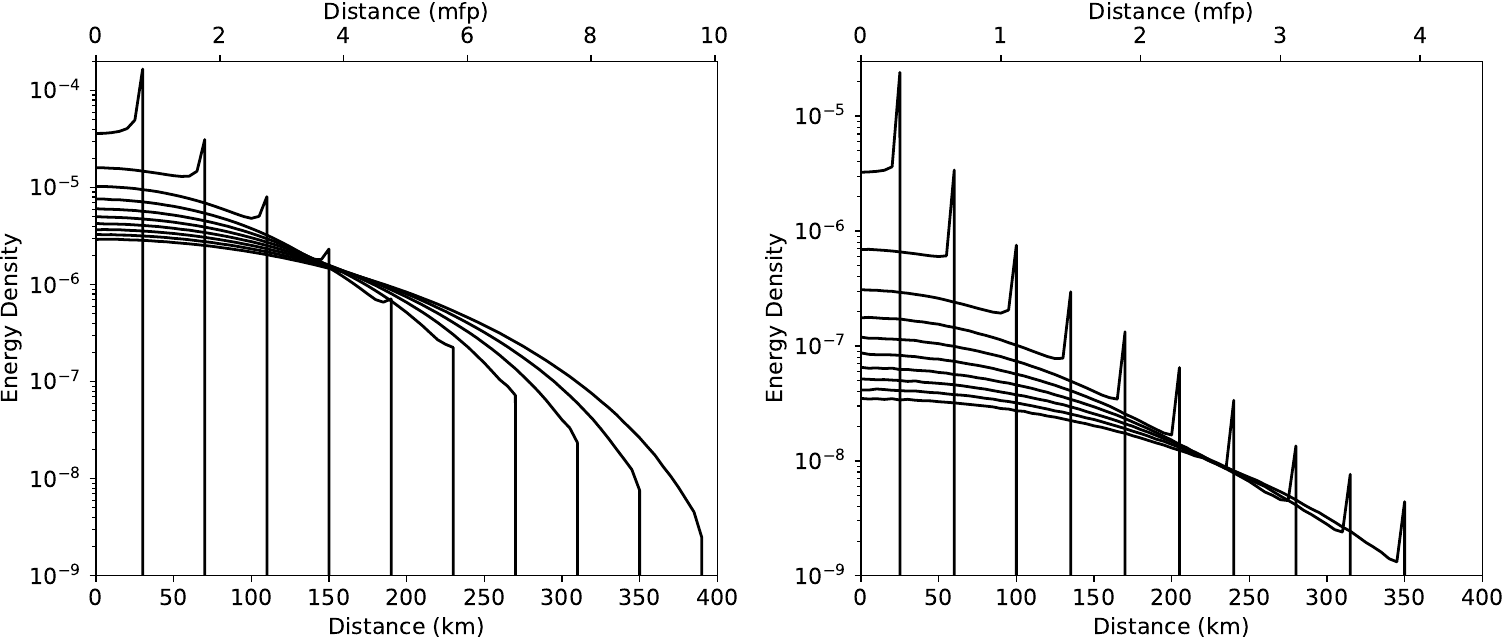}}
 \caption{  Same as Figure \ref{snapshallow} but the coupling between surface and body waves has been deactivated. See text for further explanations.  }\label{snapshallownocoupling}
 \end{figure}
 \begin{figure}
\centerline{\includegraphics[width=0.94\linewidth]{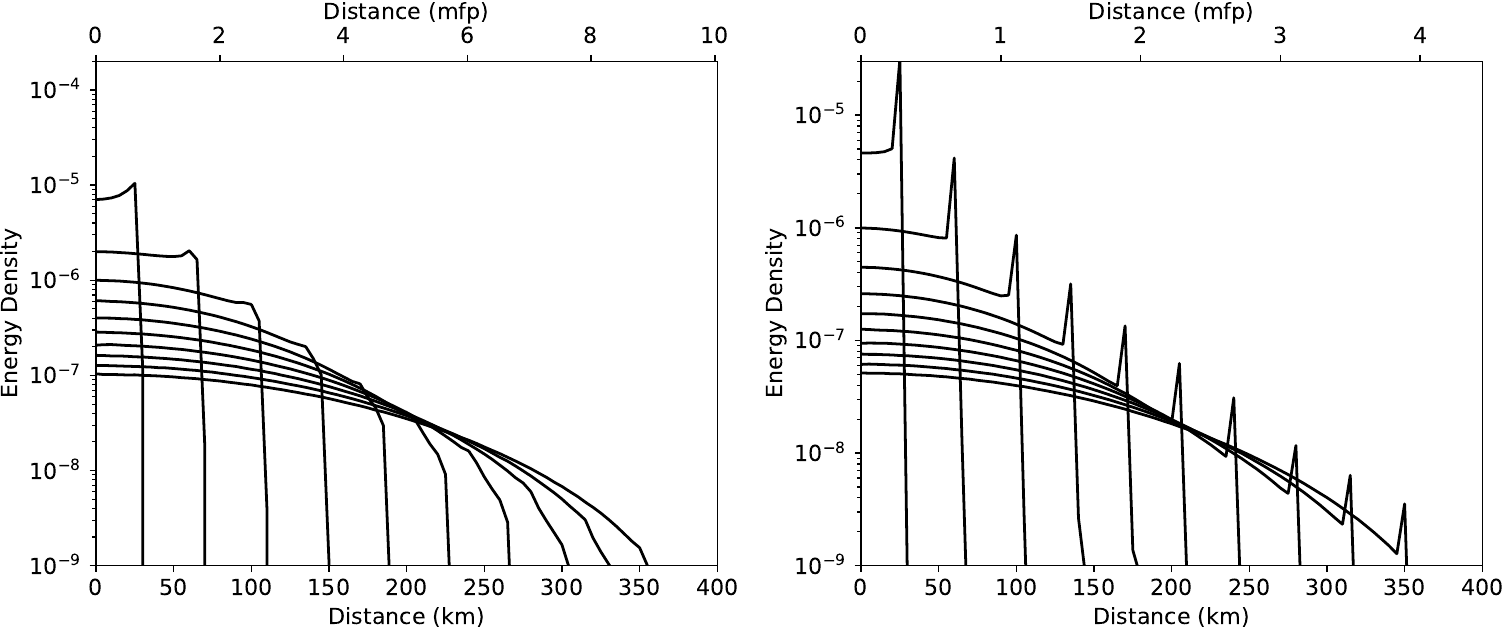}}
 \caption{Snapshots of the energy density of surface waves $\epsilon_s$ (left) and of the volumetric energy density of body waves $E_b$ (right) at the surface of a heterogeneous half-space filled with point scatterers  in the case of a deep source ($z_0 = 5 \alpha^{-1} $).  { The horizontal axes are in units of the scattering mean free path of surface waves (left, top), body waves (right, top) and in kms (bottom). For body waves, we take the mean free path value in the bulk of the medium ($z \to \infty$).}  The energy is averaged over cylindrical shells of width $\Delta r = 5$km and depth $\Delta z =5$km.  } \label{snapdeep}
 \end{figure}
 \begin{figure}
 \centerline{\includegraphics[width=0.94\linewidth]{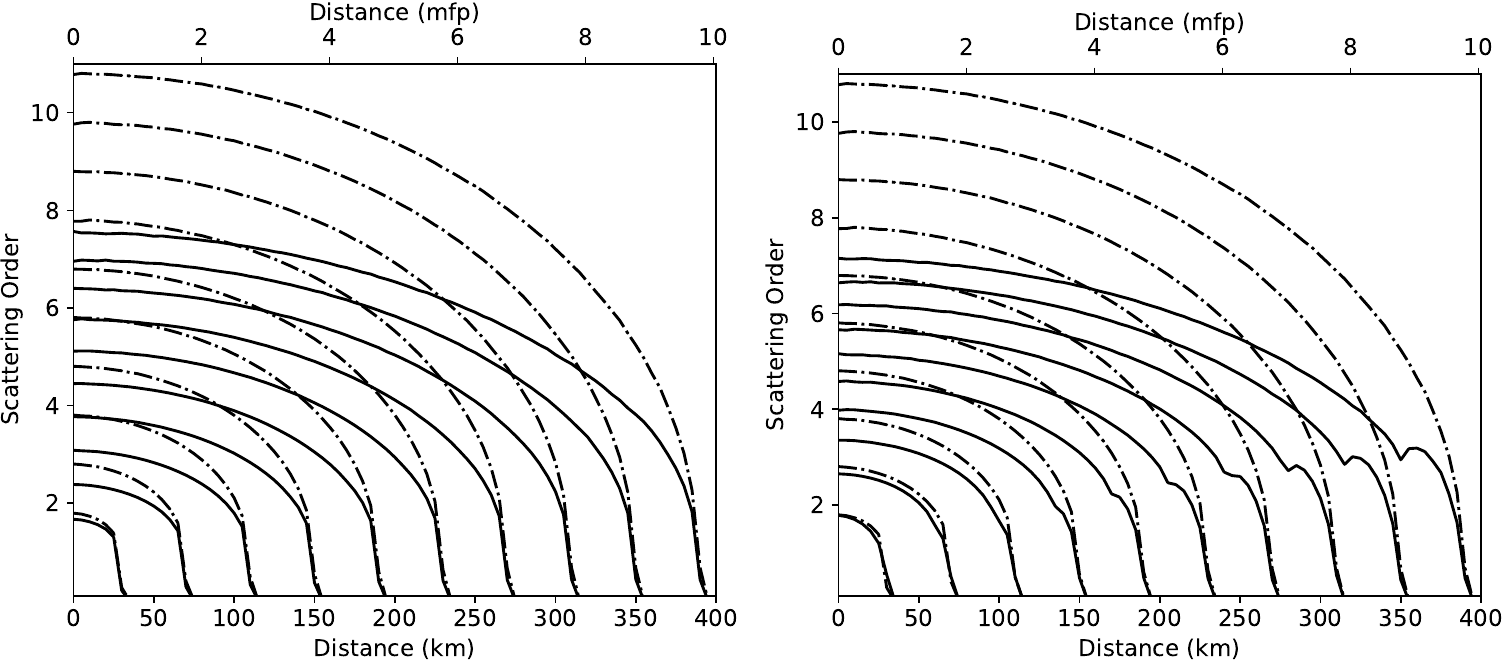}}
 \caption{ Contribution of the different orders of scattering to the energy envelopes of surface waves shown in Figure \ref{snapshallow} (left, solid lines) and \ref{snapdeep} (right, solid lines). Snapshots of the mean scattering order are represented as a function of the distance from the source. For reference, the dashed line show the same distribution for a conventional transport model in 2-D. The horizontal axes are in units of the scattering mean free path of surface waves (top) and in kms (bottom).   } \label{figorders}
 \end{figure}

In Figure \ref{snapshallow} and \ref{snapdeep}, we illustrate in greater details the multiple-scattering process by showing snapshots of the surfacic and volumetric energy densities $\varepsilon_s(t,\mathbf{r})$ and $E_b(t,\mathbf{r},z)|_{z=0}$ at regular time intervals $\Delta t=1 \tau^{s}$ starting  at a lapse time $t=0.8 \tau^{s}$ for a shallow ($z_0 =1$km) and  a deep ($z_0=5$km) source, respectively. The scattering parameters are the same as in Figure \ref{figequip} and the energy is averaged over a range of epicentral distance $\Delta r =5$km and depth $\Delta z=5$km. { We use a double horizontal axis on Figures  \ref{snapshallow}-\ref{snapdeep} to show simultaneously the epicentral distance in kms and in units of mean free path. Note that in the case of body waves, we take the value of the mean free path in the bulk of the medium $\tau^{b \rightarrow b}$.
 For comparison, we show in Figure \ref{snapshallownocoupling}, snapshots of energy density of surface waves and body waves when the coupling between the two is deactivated. In the case of surface waves, this amounts to computing the solution of a conventional 2-D multiple-scattering process with the mean free time $\tau^s$. In the case of body waves, we consider a conventional 3-D multiple-scattering process in a half-space with a constant mean free time $\tau^{b\rightarrow b}$, i.e., we remove  the boundary layer where the coupling with surface waves occurs.  To facilitate the comparison between Figure \ref{snapshallow} and  \ref{snapshallownocoupling}, we have  adjusted the  strength of the source term in the conventional multiple-scatttering simulations so that they match exactly the energy released at the source in the form of body and shear waves in the coupled case.

We first analyze the transport of surface waves in the case of a shallow source. As compared to the conventional 2-D case, 
mode coupling has at least two visible effects on the spatial distribution of the surface wave energy. First, it lowers the energy level in the coda. As an illustration, we observe that after 10 mean free times  the coda intensity is  reduced by a factor at least equal to 10 in Figure  \ref{snapshallow} compared to Figure \ref{snapshallownocoupling}.   Second, mode coupling appears to enhance the visibility of ballistic surface waves. In Figure \ref{snapshallownocoupling}, the ballistic term is completely masked by the diffuse contribution at roughly 6 mean free path from the source, whereas a small  ballistic peak is still visible at roughly 10 mean free paths from the source in Figure \ref{snapshallow}.
It is worth noting that the ballistic contribution is exactly identical in Figures  \ref{snapshallow} (left) and \ref{snapshallownocoupling} (left). Again, this is the strong decrease of the energy of scattered coda waves which explains the difference between the two Figures. Figure \ref{figorders}, which displays the spatial distribution of the mean order of scattering in the coda, reveals that the coda of coupled surface waves is depleted in high-order multiply-scattered waves compared to a conventional 2-D transport process. In other words, mode conversions entail a strong conversion of multiply-scattered  surface waves into body waves which decreases the energy level of surface wave coda and, by comparison, enhances the ballistic contribution.
Examination of Figures  \ref{snapshallow} (right) and \ref{snapshallownocoupling} (right) reveals that the effects of mode coupling on body waves are opposite to the ones just described for surface waves. Thus, we observe that the energy level in the coda is slightly increased by the transfer energy from surface wave to body waves. An additional contribution comes from the increase of the scattering strength of body waves near the surface which attenuates the ballistic waves and transfers their energy into the coda. Examination of the decay of the ballistic peak of body waves with epicentral distance in Figures  \ref{snapshallow} and  \ref{snapshallownocoupling} confirms the increased attenuation entailed by the coupling with surface waves.  Other more exotic phenomena are also visible in Figure \ref{snapshallow} such as some precursory body waves arrival due to the coupling from  surface waves  to body waves. However this process is a very peculiar feature of our model, due to the higher wavespeed of surface waves compared to the one of body waves.


Further  differences between our coupled model for surface and body waves and conventional transport theory is illustrated in Figure \ref{snapdeep} where we show snapshots of the energy distribution of surface and body waves in the case of a deep source. Note that in that case, surface waves can only be generated by mode conversions so that ballistic arrivals are absent in Figure \ref{snapdeep} (left). Interestingly, our numerical simulations indicate that surface waves are  rapidly excited to a non-negligible level in the coda.  Examination of Figure \ref{figorders} (right)  further indicates that  multiple-scattering is at the origin of the generation of surface waves in the coda when the source is located at large depth. These observations agree with our theoretical analysis of equipartition, which implies that, independent of the source depth, the coda at the surface of a half-space always appears as a mixture of surface and body waves. 
 }

  \begin{figure}
\centerline{\includegraphics[width=0.98\linewidth]{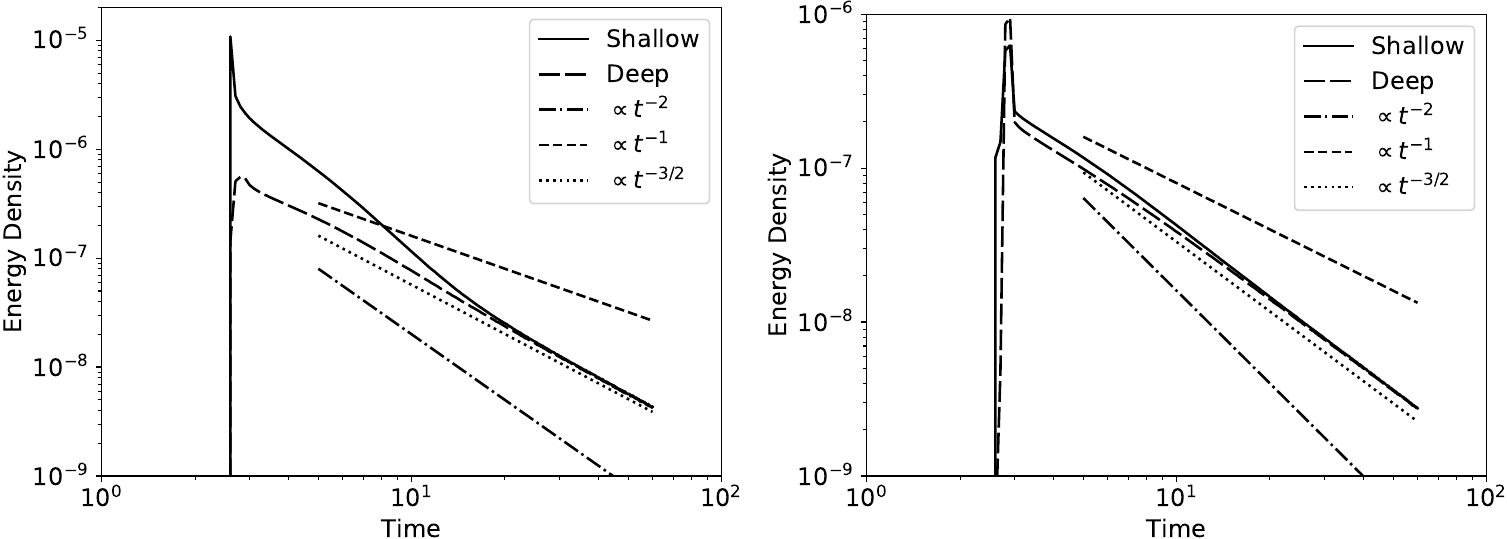}}
 \caption{Energy density of body waves $E_b$ (left) and surface waves $\epsilon_s$ (right) at the surface of a heterogeneous half-space filled with point scatterers in the case of a shallow source ($z_0 = \alpha^{-1} $). 
 The enegy is averaged over a depth $\Delta z=5$km and an epicentral distance range $\Delta r=5$km. The station is located at an epicentral distance of 50km. The horizontal axis is in units of the scattering mean free time of surface waves in logarithmic scale. Typical algebraic decays  are also shown. } \label{envelopes}
 \end{figure}

In Figure \ref{envelopes}, we show envelopes of energy densities for surface and body waves in the case of a shallow source ($z_0=1$km) { and a deep source ($z_0=5$km)} at an epicentral distance of 50km. The scattering parameters are the same as in all previous Figures and the spatial resolution of the computation is 5km again. {  The impact of the depth of the source on the excitation of  ballistic waves is obvious in Figure \ref{envelopes} and confirms the analysis of Figures \ref{snapshallow} and \ref{snapdeep}. In particular,  it is apparent that the direct body waves are less attenuated in the case of the deep source, as a consequence of the exponential decay of the scattering conversions from body to surface waves with depth.}
To facilitate the identification of different propagation regimes in the coda, we have superposed on the graphs some typical algebraic decays: $t^{-1}$ for scattering in 2-D \citep[either single or multiple, see e.g.][]{paasschens1997}, $t^{-3/2}$ for multiple scattering in 3-D, and $t^{-2}$ for single-scattering in 3-D. In Figure \ref{envelopes}, we observe that for both body and surface waves the coda obeys a $t^{-3/2}$  decay law at long lapse-time, independent of the source depth, which is characteristic of a 3-D diffusion process. This supports the predictions of  the diffusion model and confirms the dominance of body waves in the transport process at large lapse-time. 
{ At short lapse-time, we observe a distinct behavior between the two kinds of waves, particularly in the case of a shallow source}. After the passage of the ballistic waves, body waves appear to decay slightly more slowly than the asymptotic $t^{-3/2}$ behavior. This may reflect the conversion of surface waves to body waves as discussed in the analysis of Figure \ref{snapshallow}. Two propagation regimes show up clearly on the surface wave energy envelope, with a transition between the two around a lapse-time of 10 mean free times. At short time, the decay of surface waves appears to be faster than the one of body waves, probably as a consequence of the transfer of surface wave energy into the volume as discussed in relation with Figure \ref{snapshallow}. Taken together, Figures \ref{snapshallow}-\ref{envelopes} illustrate the much richer behavior of the coupled system of transport Eq. (\ref{rtb}), compared to the  {  conventional transport process without coupling between surface and body waves.}
 \section{Conclusions}
 This work represents a first attempt at formulating a self-consistent theory of RT of seismic waves in a half-space geometry including the coupling between surface and body waves. The main approximation underlying our work is that, upon reflection at the surface, the phase of body waves is randomized so that upgoing and downgoing fluxes may be considered as statistically independent. { Our approach distinguishes itself from the standard Eqs of RT for scalar waves found in the literature in one important way: it keeps track - to some extent - of the wave behavior in the vicinity of the surface. This has a number of consequences: (1) surface and body waves are coupled by conversion scattering (2) even in a statistically homogeneous medium it requires that the scattering properties of body waves depend on the depth in the medium.}  Furthermore, { the} reciprocity relation between the surface-to-body vs body-to-surface mean free times plays a prominent role in the establishment of an equipartition regime with a ratio that conforms to the  predictions of standard mode counting arguments. Besides equipartition, a notable outcome of our RT equations is the anisotropy of the diffusivity  of seismic waves, due to the difference in scattering properties and wave velocities of body and surface waves. We also show that our RT Eqs are operational, in the sense that they are readily amenable to numerical solutions by Monte-Carlo simulations. These simulations could be used in the future to study in more details the dynamics of equipartition, in particular, how the equipartition time varies as a function of the ratio between the penetration depth of surface waves and the scattering mean free path for body-to-surface wave coupling.

Before becoming a viable alternative to current approaches, our theory needs to be tested and improved. In the future, we plan to address the following issues: (1) Evaluate the impact of neglecting the interference between upgoing and downgoing body waves on the scattering cross-section and, if possible, go beyond this approximation.  (2) Extend the theory to more realistic finite size scatterers and more general spatial distributions of scatterers. (3) Incorporate polarization effects for elastic waves at a free surface.  { (4) Absorption of energy is also a very important mechanism of attenuation, which has been entirely neglected in this work  for simplicity. Because the sub-surface of the Earth is thought to be very strongly attenuating due to the widespread presence of fluids, we may expect dissipation to affect more severely surface waves  than body waves. In turn, this may modify the partitioning of the energy in the coda as was previously shown by \citet{margerin2001} in the case of coupled $S$  and $P$   waves. Special efforts should be devoted to this important topic  before our formalism can be applied to real seismic data.}  
\begin{acknowledgments}
{The authors wish to thank the Associate Editor S. Ni and an anonymous referee for their suggestions to clarify the presentation of the results. The careful comments and constructive criticisms of H. Sato contributed to significant changes and improvements in the content of the manuscript. The authors acknowledge  the European Research Council under the European Union Horizon 2020 research and innovation program (grant agreement no. 742335 - F-IMAGE). }
\end{acknowledgments}


 \appendix
\section{Variational formulation for mixed boundary conditions} \label{variational}
{ Here, we recall briefly on a simple one-dimensional example how mixed boundary conditions of the type used in Eq. (\ref{BC}) can be incorporated into a variational formulation. The interested reader will find further details and more examples in the classic book by \citet{gelfand1963}, after which our treatment is modeled.  For simplicity we consider a vibrating string of  density $\rho$, tension $T$ and length $L$. For the moment, we do not specify the boundary conditions.  The total kinetic energy stored in the string at time $t$ is given by:
\begin{equation}
 T[u](t) = \int_0^L \rho (\partial_t u(\mathbf{x},t))^2 dx,
\end{equation}
where $u$ denotes the displacement field. The instantaneous potential energy stored in the string may be expressed as
\begin{equation}
 V[u](t) = \int_0^L T (\partial_x u(\mathbf{x},t))^2 dx \label{v0}
\end{equation}
According to Hamilton's principle, among all possible displacement fields, the one that satisfies the actual equations of motion  should make the following action integral: \begin{equation}
 I[u] = \int^{t_2}_{t_1} (T - V)[u](t)  dt
\end{equation}
stationary.
Mathematically, this principle of stationary action may be expressed as:
\begin{equation}
\partial_{\epsilon} I[u + \epsilon \psi]|_{\epsilon=0} = 0
\end{equation}
where $\psi$ is an arbitrary function. This is sometimes written as $\delta I =0$, where $\delta I$ is known as the first variation of the action integral.
Using integration by parts, \citet{gelfand1963} establish that:
\begin{equation}
\begin{split} 
\delta I = &\epsilon \left( \int^{t_1}_{t_0} \int_0^L  ( -\rho \partial_{tt} u(\mathbf{x},t) +T \partial_{xx} u(\mathbf{x},t)) \psi(x,t) dxdt  \right. \\
                 &+ \left. T \int_{t_0}^{t_1} (\partial_x u(0,t)\psi(0,t)   -\partial_x u(l,t)\psi(l,t))dt    \right) \\ \label{deltai}
\end{split}                 
\end{equation}
The arbitrariness of the function $\psi$ in Eq. (\ref{deltai}) implies  both the governing wave equation  for the vibrating string:
\begin{equation}
\rho \partial_{tt} u(x,t) - T \partial_{xx} u(x,t) =0
\end{equation}
as well as the so-called natural boundary conditions:
\begin{equation}
 \partial_x u(x,t)|_{x=0} = 0 \text{ and }  \partial_x u(x,t)|_{x=l} = 0,
 \end{equation}
 which correspond to a string with free ends. In order to obtain mixed boundary conditions, it suffices to add to the potential energy (\ref{v0}), a term of the form 
 $ \chi u(0,t)^2 $ where $\chi$ is a constant. Eq. (\ref{deltai}) must be modified accordingly by adding the new contribution  $-\epsilon \chi \int_{t_0}^{t_1} u(0,t) \psi(0,t) dt $ which, in turn, implies a natural boundary condition of the mixed type at $x=0$:
 \begin{equation}
 \left( \chi \partial_x u(x,t)  -T u(x,t) \right)|_{x=0}= 0
 \end{equation}
  The total potential energy  may be rewritten in integral form  as follows:
 \begin{equation}
 V[u](t) = \int_0^L  \left[ \chi u(0,t)^2\delta(x)  + T (\partial_x u(\mathbf{x},t))^2 \right] dx 
\end{equation}
 which justifies the appearance of the delta function in Eq. (\ref{lagrangian}) and Eq. (\ref{w}) in a simplified context.}
 \section{Far-field expression of the Green's function for scalar waves in a half-space with mixed B.C.} \label{gfderiv}
 { In this Appendix, we summarize the key steps to the derivation of Eq. (\ref{gff}  ) from Eq. (\ref{gex}). We split the computation into two parts and begin with the surface wave contribution:
 \begin{equation}
G_s(\mathbf{r},z,z_0) = \dfrac{2 \alpha e^{-\alpha(z+z_0)}}{(2 \pi)^2}    \int_{\mathbb{R}^2} \dfrac{e^{i \mathbf{k}_{\parallel} \cdot \mathbf{r} } }{k^2 +\alpha^2 -k_{\parallel}^2 +i\epsilon}    d^2k_{\parallel} 
 \end{equation}
 Introducing cylindrical coordinates $(k_{\parallel},\phi)$ and integrating over angle yields:
 \begin{equation}
G_s(\mathbf{r},z,z_0) =  \dfrac{2 \alpha e^{-\alpha(z+z_0)}}{2 \pi}    \int_0^{+\infty} \dfrac{J_0(k_{\parallel} r) }{k^2 +\alpha^2 -k_{\parallel}^2 +i\epsilon}    dk_{\parallel} 
\end{equation}
where $J_0$ denotes the standard Bessel function of order 0.
Using the same trick as in \citet[][Chapter 6]{akirichards}, we extend the wavenumber integral over the whole $k_{\parallel}$ axis using the Hankel function of the first kind instead of the Bessel function:
\begin{equation}
G_s(\mathbf{r},z,z_0) =\dfrac{ \alpha e^{-\alpha(z+z_0)}}{2 \pi}    \int_{-\infty}^{+\infty}  \dfrac{H^{(1)}_0(k_{\parallel} r) }{k^2 +\alpha^2 -k_{\parallel}^2 +i\epsilon} dk_{\parallel}
\end{equation}
In the last step, we employ the  residue theorem by closing the contour in the upper half of the complex plane with a semi-circle of radius $R \to +\infty$ and note the presence of pole at $k_{\parallel} = \sqrt{k^2 + \alpha^2} +i\eta$, $(\eta \to 0^{+})$. Thanks to the exponential decay of the integrand, the integral on the semi-circle vanishes which yields:
\begin{equation}
G_s(\mathbf{r},z,z_0) =\dfrac{ -i \alpha }{2 \pi}e^{-\alpha(z+z_0)} H^{(1)}_0( \sqrt{k^2 + \alpha^2} r).  \label{gexact}
\end{equation}
The result (\ref{gexact}) is exact. The far-field approximation (\ref{gff}) follows by application of standard asymptotic expansions to the Hankel function.

The computation of the body wave contribution can also be split into two parts:
\begin{equation}
\begin{split}
G_b(\mathbf{r},z,z_0) = & \dfrac{1}{(2\pi)^3} \int^{+\infty}_0 dq \int_{\mathbb{R}^2} \dfrac{e^{i \mathbf{k}_{\parallel} \cdot \mathbf{r} } (e^{-i q z}  +r(q) e^{iq z}) (e^{-i q z_0}  +r(q) e^{iq z_0})^* }{k^2 - k_{\parallel}^2 -q^2 +i \epsilon } d^2k_{\parallel}  \\
                  = & \dfrac{1}{(2\pi)^3} \int^{+\infty}_{-\infty} dq  \int_{\mathbb{R}^2} \dfrac{e^{iq(z-z_0)}    e^{i \mathbf{k}_{\parallel} \cdot \mathbf{r} }}{k^2 - k_{\parallel}^2 -q^2 +i \epsilon } d^2k_{\parallel}   \\
                    & +   \dfrac{1}{(2\pi)^3} \int^{+\infty}_{-\infty} dq  \int_{\mathbb{R}^2} \dfrac{r(q) e^{iq(z+z_0)}   e^{i \mathbf{k}_{\parallel} \cdot \mathbf{r} }}{k^2 - k_{\parallel}^2 -q^2 +i \epsilon } d^2k_{\parallel} \\
                    =& G_{\infty}(\mathbf{r},z,z_0) + G_{r}(\mathbf{r},z,z_0)  \label{gb}
\end{split}
\end{equation}
where the unitarity of the reflection coefficient has been used and the $q$ integral has been extended from $-\infty$ to $+\infty$ thanks to the relation $r(q)^*=r(-q)$.
The first term in the second equality of (\ref{gb}) may be recognized as the full-space solution to the Helmholtz Eq.:
\begin{equation}
\begin{split}
 G_{\infty}(\mathbf{r},z,z_0) = & \dfrac{1}{(2\pi)^3} \int^{+\infty}_{-\infty} dq  \int_{\mathbb{R}^2} \dfrac{e^{iq(z-z_0)}    e^{i \mathbf{k}_{\parallel} \cdot \mathbf{r} }}{k^2 - k_{\parallel}^2 -q^2 +i \epsilon } d^2k_{\parallel} \\
              =&  -\dfrac{e^{ikR_0}}{4 \pi R_0},
 \end{split}
\end{equation}
where $R_0 = \sqrt{r^2 + (z -z_0)^2}$. The second  term in the second equality of (\ref{gb}) represents the waves reflected at the surface:
\begin{equation}
 G_{r}(\mathbf{r},z,z_0) =  \dfrac{1}{(2\pi)^3} \int^{+\infty}_{-\infty} dq  \int_{\mathbb{R}^2} \dfrac{r(q) e^{iq(z+z_0)}   e^{i \mathbf{k}_{\parallel} \cdot \mathbf{r} }}{k^2 - k_{\parallel}^2 -q^2 +i \epsilon } d^2k_{\parallel} 
\end{equation}
 The computation of this integral  may be attacked in exactly the same way as we did for the surface wave term $G_s$ to obtain:
\begin{equation}
 G_{r}(\mathbf{r},z,z_0) =  \dfrac{-i}{8\pi} \int^{+\infty}_{-\infty}   r(q) e^{iq(z+z_0)}  H^{(1)}_0( \sqrt{k^2 -q^2} r)dq,
\end{equation}
To approximate this last integral in the far-field of the source, we first remark that for $|q| > k$ the cylindrical waves are evanescent so that we may legitimately take $-k$ and $+k$ as integration limits. We next make use of the far-field expansion of the Hankel function to obtain the following oscillatory integral representation:
\begin{equation}
 G_{r}(\mathbf{r},z,z_0) \approx  \dfrac{-i}{8\pi} \sqrt{\dfrac{2}{\pi r}}  \int^{+k}_{-k}   \dfrac{ r(q) e^{iq(z+z_0)   +i\sqrt{k^2 -q^2}r -i\pi/4} }{\sqrt{k^2 -q^2} }dq
\end{equation}
 Further noting that the derivative of  the phase term:
\begin{equation}
    \phi(q) = q (z + z_0) + \sqrt{k^2-q^2} r
\end{equation}
vanishes at :
\begin{equation}
q_0= \dfrac{k(z+z_0)}{R_0'}   \label{q0}
\end{equation}
 with $R'_0= \sqrt{r^2 + (z+z_0)^2}$,  we apply the stationary phase formula to obtain after some straightforward  algebra:
\begin{equation}
 G_{r}(\mathbf{r},z,z_0) \approx  - \dfrac{r(q_0) e^{i k R'_0}}{4 \pi R'_0}. \label{gref}
\end{equation}
This term may be interpreted as the contribution of the image point of the source with a strength given by the reflection coefficient evaluated at an incidence angle corresponding to the specularly reflected ray connecting the source to the detection point (see Eq. \ref{q0}). To complete the far-field approximation, we first note the following expansions: $R_0 = R -z_0z/R + o(1/R)$, $R'_0=R+z_0z/R +o(1/R)$ where $R=\sqrt{r^2 + z^2}$. Neglecting all terms smaller than $1/R$ for the amplitude, all terms smaller than $z_0/R$ for the phase and further approximating $q_0$ as $kz/R$, formula (\ref{gff}) is recovered.
 }
 
 \end{document}